%% file: long.tex
\documentclass[pra,amsmath,superscriptaddress,twocolumn]{revtex4-2}

\usepackage{epsfig}
\usepackage{hyperref}
\usepackage{amssymb}
\usepackage{amsmath}
\usepackage{amsfonts}
\usepackage{soul}    
\usepackage{bm}
\usepackage{braket}
\usepackage{graphicx}
\usepackage{xcolor}
\usepackage{float}
\usepackage{lipsum}
\usepackage{comment}
\usepackage{siunitx}

\renewcommand{\d}{\ensuremath{\mathrm{d}}}
\newcommand{\e}{\ensuremath{\mathrm{e}}}

\newcommand{\abs}[1]{\left\lvert#1\right\rvert} 
\newcommand{\kf}{k_{\rm{F}}}
\newcommand{\kp}{k_{\rm{p}}}
\newcommand{\kh}{k_{\rm{h}}}
\newcommand{\mc}[1]{\mathcal{#1}}
\newcommand{\rhoi}{\rho_{\rm{i}}}
\newcommand{\bef}{\ensuremath{b}} 

\include{./journalsVaps}

\begin{document}
\title{Finding the phase diagram of strongly-correlated disordered bosons using quantum quenches} 
\author{L. Villa}
\email{louis.villa@polytechnique.edu (he/him/his)}
\affiliation{CPHT, CNRS, Ecole Polytechnique, IP Paris, F-91128 Palaiseau, France}
\author{S. J. Thomson}
\email{steven.thomson@polytechnique.edu (he/him/his)}
\affiliation{CPHT, CNRS, Ecole Polytechnique, IP Paris, F-91128 Palaiseau, France}
\affiliation{JEIP, USR 3573 CNRS, Coll\`ege de France, PSL Research University, 11 Place Marcelin Berthelot, 75321 Paris Cedex 05, France}
\author{L. Sanchez-Palencia}
\email{laurent.sanchez-palencia@polytechnique.edu (he/him/his)}
\affiliation{CPHT, CNRS, Ecole Polytechnique, IP Paris, F-91128 Palaiseau, France}
\date{\today} 

\begin{abstract}
The question of how the low-energy properties of disordered quantum systems may be connected to 
exotic localization phenomena at high energy is a key open question in the context of quantum glasses and many-body localization.
In Ref.~\cite{Paper1}, we have shown that key features of the excitation spectrum of a disordered system can be efficiently probed from out-of-equilibrium dynamics following a quantum quench, providing distinctive signatures of the various phases.
Here, we extend this work by providing a more in-depth study of the behavior of the quench spectral functions associated to different observables and investigating an extended parameter regime.
We provide a detailed introduction to quench spectroscopy for disordered systems and show how spectral properties can be probed using both local operators and two-point correlation functions. 
We benchmark the technique using the one-dimensional Bose-Hubbard model in the presence of a random external potential, focusing on the low-lying excitations, and demonstrate that quench spectroscopy can distinguish the Mott insulator, superfluid, and Bose glass phases. We then explicitly reconstruct the zero-temperature phase diagram of the disordered Bose-Hubbard at fixed filling using two independent methods, both experimentally accessible via time-of-flight imaging and quantum gas microscopy respectively, and demonstrate that quench spectroscopy can give valuable insights as to the distribution of rare regions within disordered systems.
\end{abstract}

\maketitle
\section{Introduction} 

Many-body quantum systems are characterized largely by their low-lying excitations. These excitations govern a wide variety of important physical phenomena, such as transport properties and response functions, and provide a detailed insight into the nature of the system. 
Standard spectroscopic methods used to probe excitation spectra employ pump-probe techniques~\cite{damascelli2004Probing,clement2010Bragg,lewenstein2012Ultracold} based around the linear response of a system to a weak perturbation, where the `pump' excites low-lying excitations and the `probe' measures them.
In ultracold atomic gases, excitations are typically created through photon scattering by coupling to a different internal atomic state (radio frequency and Raman spectroscopy~\cite{campbell2006Imaging,stewart2008Using,dao2009Probing}) or back to the same state (Bragg spectroscopy~\cite{ernst2010Probing,clement2010Bragg,fabbri2012Quasiparticle}). The momentum of the resulting excitation is fixed by the angle between the pump lasers, which has to be fine tuned~\cite{clement2009Exploring} and thus requires tremendous work to reconstruct the excitation spectrum over the full Brillouin zone. An alternative method is lattice modulation spectroscopy, where a time-dependent modulation of the lattice potential is used to generate excitations~\cite{stoferle2004Transition,iucci2006Energy,kollath2006Spectroscopy,orso2009Lattice,gadway2011Glassy,DErrico+14,citro2020Lattice}. However this method only probes the zero-momentum sector, preventing reconstruction of the full spectrum. These traditional spectroscopic techniques are challenging to engineer experimentally, and alternative methods are desirable. Moreover, methods which rely on probing specific momentum values are not well-suited to inhomogeneous or disordered systems, where the breaking of translational invariance means that momentum is no longer a well-defined quantum number. 

New spectroscopic methods based on quantum quenches, known as quench spectroscopy, have recently emerged~\cite{gritsev2007Spectroscopy,Naldesi+16,villa2019Unraveling,villa2020Local}, where the initial `pump' step is replaced by a weak quench which populates the low-lying excited states of the model. This quench can be either global~\cite{villa2019Unraveling} or local~\cite{villa2020Local}, and is highly flexible, allowing both for the system to generate its own excitations and for careful targeting of specific types of excitations~\cite{jurcevic2015Spectroscopy}. In homogeneous systems, it has been demonstrated that spectral properties, including the dispersion relation of the elementary excitations, may then be obtained from the post-quench non-equilibrium dynamics via a straightforward Fourier transform~\cite{villa2019Unraveling,villa2020Local}.

In Ref.~\cite{Paper1}, we have developed a form of quench spectroscopy suitable for disordered systems, and which is able to distinguish between all three zero-temperature phases of the disordered Bose-Hubbard model. Moreover, we have introduced a new local probe which provides information on the spatial distribution of gapped and gapless regions, acting as an alternative probe of the various phases.
Here, we extend this work by providing a more in-depth study of the behavior of different observables and working in extended parameter regimes.
Moreover, we explicitly show that the complete phase diagram of the disordered Bose-Hubbard model can be obtained from quench spectroscopy.

\begin{center}
\begin{figure}[t!]
\includegraphics[width= \linewidth]{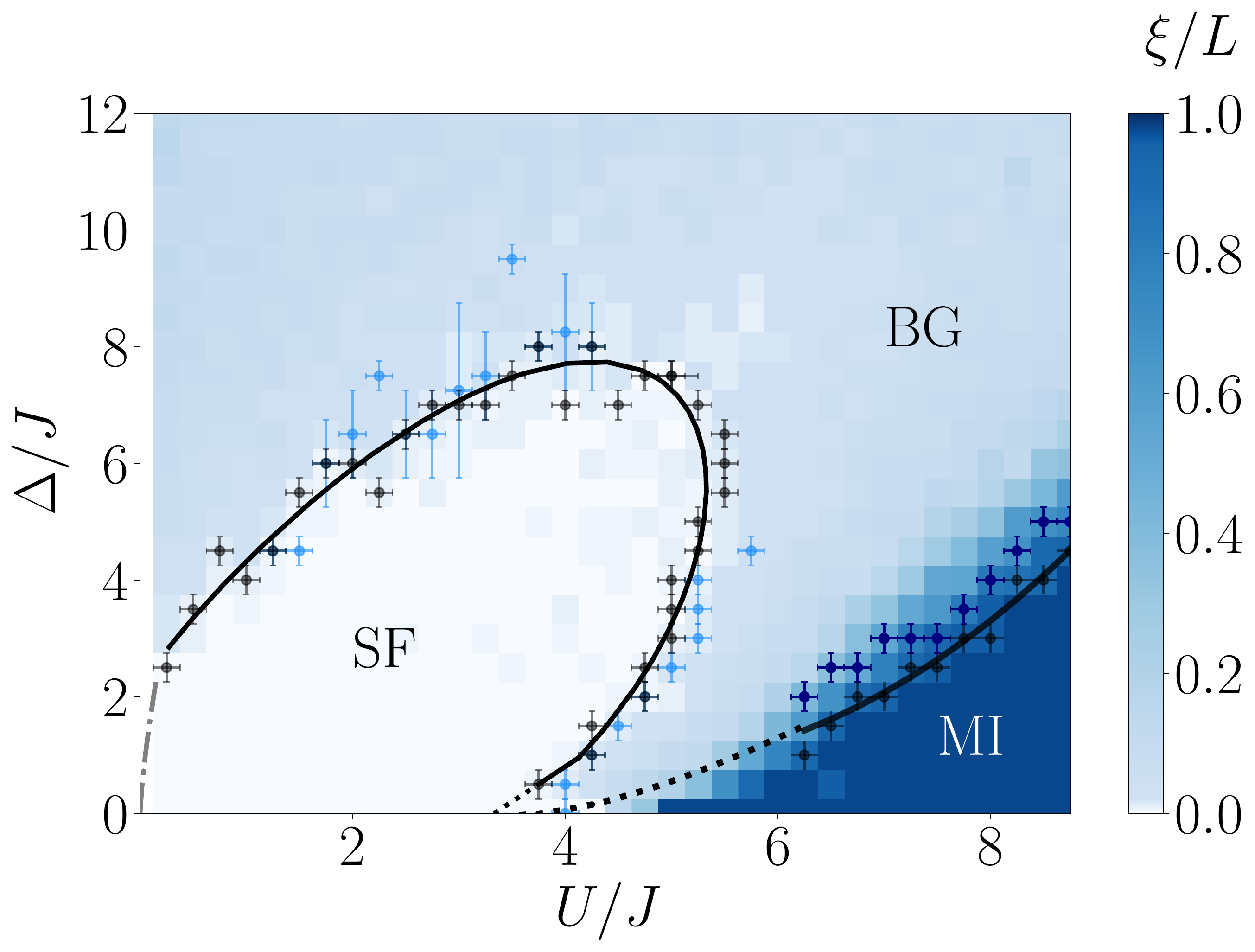}
\caption{
A summary of our results reconstructing the phase diagram of the disordered Bose-Hubbard chain at fixed filling $\overline{n}=1$ with a random on-site potential of strength $\Delta/J$, obtained using quench spectroscopy (system size $L=47$, averaged over $N_{\rm{s}}=25$ disorder realizations). 
The black points are obtained from analysis of the local spectral function (LSF, see Sec.~\ref{sec.local}), which provides information about the real-space distribution of gapped and gapless regions; the solid black line is a guide-to-the-eye fit of these points. 
The blue points are obtained from the excitation spectrum, which allows us to identify the SF-BG transition based on where the sound velocity drops to zero (light blue) and the MI-BG transition based on where the single-particle energy gap closes (dark blue). It yields a fair estimate of the transitions for moderate disorder, although less accurate than the LSF approach.
The dotted black line is an approximation to the MI-BG phase boundary obtained in a region where the gap is smaller than our numerical resolution, see Sec.\ref{sec.phase} for details. 
The grey line close to the origin is a power-law of the form $\Delta \propto U^{3/4}$~\cite{lugan2007a,pollet2013Review}. 
}
\label{fig.phase}
\end{figure}
\end{center}

By measuring the post-quench dynamics of equal-time observables, we demonstrate how quench spectroscopy grants momentum-resolved information about the excitations in the context of disordered ultracold atomic gases and trapped ion systems. Both one-point functions and equal-time correlators are advantageously accessible through experimental snapshots, for instance through single-atom resolution imaging using quantum gas microscopes~\cite{Bakr+09,Sherson+10,Bakr+10,Haller+15} or after a time-of-flight measurement~\cite{lewenstein2012Ultracold,bloch2008Manybody}. 
We emphasize the essential differences between homogeneous and disordered quantum matter from the point of view of quench spectroscopy, and compare with numerically exact tensor network simulations.
We show that this method allows direct measurement of Griffiths effects in disordered systems, providing an alternative way in which the phase diagram can be constructed from the spatial distribution of locally gapped and gapless regions within a disordered system.
The final result of this procedure is shown in Fig.~\ref{fig.phase}, where we plot the zero-temperature phase diagram of the disordered Bose-Hubbard chain at fixed filling $\overline{n}=1$ using the typical size of locally gapped regions as the order parameter. This allows us to clearly distinguish regions of the phase diagram in which all sites host gapless excitations (superfluid), all sites host gapped excitations (Mott insulator), and the intermediate case (Bose glass). The remainder of this paper will discuss the details of how this phase diagram is computed, and demonstrate the advantages of the quench spectroscopy method.

The paper is organized as follows. In Sec.~\ref{sec.model}, we summarize the disordered one-dimensional Bose-Hubbard model which we shall use for our analysis. In Sec.~\ref{sec.qsf} we discuss the application of quench spectroscopy to disordered quantum systems and discuss the role played by disorder.
In Sec.~\ref{sec.strong_int} we present numerical results for the disordered Bose-Hubbard chain, comparing several observables and expanding upon the results of Ref.~\cite{Paper1}. We discuss how different observables probe distinct spectral properties, which can be used to describe various features of the elementary excitations within each phase, and enables direct measurement of the speed of sound in the superfluid phase. We then investigate how the spectral properties are modified by increasing the disorder strength at fixed unit filling in Sec.~\ref{sec.results2}. In Sec.~\ref{sec.local} we introduce a site-resolved local spectral function and demonstrate that this gives a particularly convenient method to reconstruct the full phase diagram, as shown in Fig.~\ref{fig.phase}. 
Finally we conclude in Sec.~\ref{sec.conclusion} with a discussion and outlook for future work.

\section{The disordered Bose-Hubbard model}
\label{sec.model}
We shall focus on one-dimensional ultracold atomic gases, where disorder has already been investigated~\cite{sanchez-palencia2010Disordered} in the contexts of Anderson localization~\cite{Billy+08,Roati+08,Kondov+11,Jendrzejewski+12,Semeghini+15},
collective localization~\cite{gurarie2002,gurarie2003,bilas2006,lugan2007b,lugan2011,lellouch2015},
many-body localization~\cite{Schreiber+15,Bordia+16,Choi+16} and quantum glasses~\cite{Fallani+07,White+09,Deissler+10,Pasienski+10,Gadway+11,DErrico+14,Meldgin+16}. 
The behavior of ultracold bosons in a one-dimensional lattice is described by the Bose-Hubbard model, which in the presence of disorder is given by
\begin{equation}
\label{eq:hamiltonian_bhmd}
\begin{split}
\hat{H}=\sum_{j}\left[-J\left(\hat{a}_{j}^{\dagger}\hat{a}_{j+1}+\text{h.c.}\right)+\frac{U}{2} \hat{n}_j (\hat{n}_j - 1) + \mu_j\hat{n}_{j}\right],
\end{split}
\end{equation}
where $\hat{a}_{j}^{\dagger}$ and $\hat{a}_{j}$ are respectively the creation and annihilation operators of a boson on the lattice site $j$ and $\hat{n}_j=\hat{a}^{\dagger}_{j}\hat{a}_{j}$ the associated density. The lattice spacing is set to one throughout the following. The disorder is contained in the term $\mu_{j}$, given by $\mu_j=\Delta_j - \mu$, with $\mu$ the chemical potential, in the grand canonical ensemble, or by $\mu_j=\Delta_j$ in the canonical ensemble where the number of particles is fixed. We shall consider both cases in the following. We will consider random disorder drawn from a box distribution, with $\Delta_j \in [-\Delta/2,\Delta/2]$ where $\Delta$ parametrizes the overall disorder strength.
This type of disorder can be approximated by speckle patterns~\cite{clement2005Suppression,clement2006Experimental,Billy+08,White+09} or created exactly by a spatial light modulator \cite{Choi+16,Bruce+15}. 

\begin{center}
\begin{figure}[t!]
\includegraphics[width=0.9\linewidth]{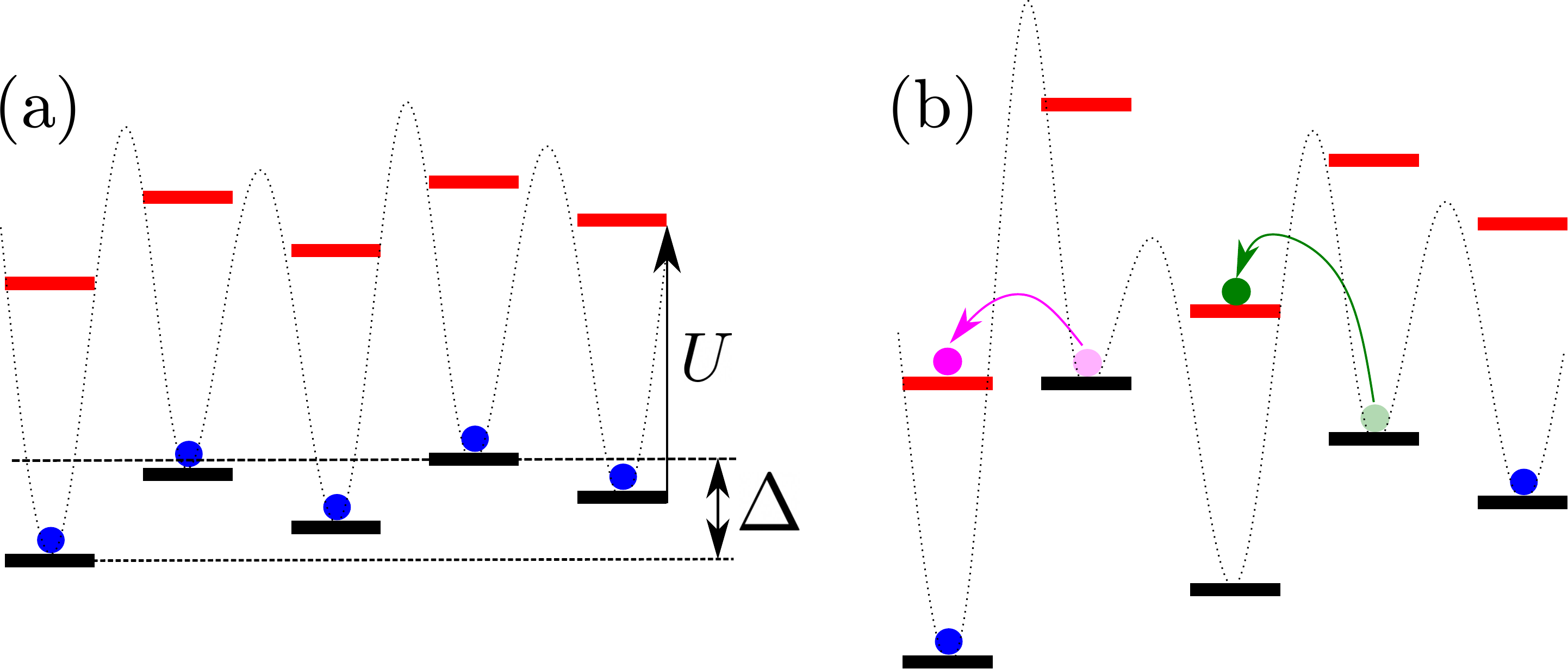}
\caption{Sketch of local excitations in the atomic limit with $\bar{n}=1$. Each site is represented by a two-level system (black, red) separated by an energy $U$. The circles represent bosons occupying some energy levels, and the dashed lines show the disordered potential, creating a random modulation of the on-site energies. (a)~MI in the weak disorder regime $\Delta<U$ where all local excitations are gapped. (b)~BG in the strong disorder regime (here for $\Delta=U$) showing the coexistence of locally gapped (green) and gapless (magenta) excitations. The curved arrows in corresponding colors represent the displacement of atoms generating such excitations.}
\label{fig.sketch}
\end{figure}
\end{center}

The disordered Bose-Hubbard model at zero temperature contains three phases.
The Mott insulator (MI), where strong interactions prevent particle transport, is characterized by its incompressible nature and gapped excitations.
The superfluid (SF), by contrast, is gapless and compressible, exhibiting quasi-long-range order in one dimension.
In the presence of a random external potential, a Bose glass (BG) phase~\cite{Giamarchi+88,Fisher+89} intervenes between the SF and the MI. The BG is a gapless, compressible insulator, and can host coexistinglocal MI (gapped) and SF (gapless) regions within a single sample, as sketched in Fig.~\ref{fig.sketch}.

Despite intense analytical~\cite{Mukhopadhyay+96,Freericks+96,Svistunov96,Herbut97,Herbut98,Buonsante+07,Weichman+08,Bissbort+09,Kruger+09,Pollet+09,Bissbort+10,Kruger+11,Stasinska+12,Hegg+13,Thomson+14,Thomson+15,Dupuis19} and numerical~\cite{Scalettar+91,Krauth+91,Kisker+97,Sen+01,Lee+01,Prokof'ev+04,Gurarie+09,Niederle+13,alvarezzuniga2013Boseglass,Doggen+17,yao2020LiebLiniger,gautier2021Strongly} study, many questions as to the nature of the BG remain.
Various theoretical proposals have been put forward suggesting ways to observe the BG in experiments~\cite{Morrison+08,roscilde2009Probing,delande2009Compression,roux2013Dynamic,Thomson+14,Thomson+16,yao2020LiebLiniger,gautier2021Strongly}.
However as the BG does not break any physical symmetries, it has proven highly challenging to detect. Several experimental approaches have been employed to date, such as in optical lattices using time-of-flight imaging~\cite{Pasienski+10,Meldgin+16} and Bragg spectroscopy~\cite{Fallani+07}, as well as thermodynamic measurements~\cite{Yamada+11,yu2012Bose} and neutron scattering~\cite{Hong+10} in solid state magnets~\cite{Zheludev+13}. Most standard techniques - with a few exceptions~\cite{Morrison+08,Thomson+16} - rely on a comparison of several global probes in order to distinguish the BG from the SF and MI phases.
In contrast, in this work, we present two independent probes able to uniquely identify the BG.

\section{Quench Spectroscopy in the presence of disorder}
\label{sec.qsf}
\subsection{General strategy}
The general idea behind quench spectroscopy in ultracold atomic gas platforms is to replace the `pump' step used in typical pump-probe techniques with a quantum quench, which generates excitations in the system that can then be measured using standard imaging techniques.
The strength of the quench plays a similar role as the initial temperature in other spectroscopic methods (see Appendix~\ref{app.temp} for details). Spectral information can be obtained from the resulting non-equilibrium dynamics via a straightforward Fourier transform. Quenches have previously been used to identify quantum phase transitions~\cite{Bhattacharyya+15,Heyl+18,Titum+19,Haldar+20}. 
The main object of interest in quench spectroscopy is the \emph{quench spectral function} (QSF). After a quench, the out-of-equilibrium dynamics of a given observable $\hat{O}(x,t)$ are given by
\begin{equation}
\label{eq:observable}
G(x,t)=\langle\hat{O}(x,t)\rangle=\text{Tr}\left[\hat{\rho}_{\textrm{i}}\,\hat{O}(x,t)\right],
\end{equation}
where $\hat{\rho}_{\rm{i}}$ is the density matrix of the initial state.
The QSF is then obtained from a space-time Fourier transform of Eq.~\eqref{eq:observable}, leading to
\begin{equation}
\label{eq:QSF_disorder_full_expression}
\begin{split}
G(k,\omega)&=2\pi\sum_{\nu,\nu'}\rho_{\rm{i}}^{\nu'\nu}\,\delta(E_{\nu'}-E_{\nu}-\omega)\\
&\qquad\times\int\d x\,\e^{-ikx} \bra{\nu} \hat{O}(x)\ket{\nu'},
\end{split}
\end{equation}
where $\{\ket{\nu}\}$ denotes the eigenstates of the post-quench Hamiltonian for a fixed disorder realization. 

For weak quenches, the initial state is close to the ground state of the post-quench Hamiltonian, and $\rho_{\rm{i}}^{\nu'\nu}$ is significant only for $\nu=0$ or $\nu'=0$. Restricting the discussion to the positive frequency sector ($\omega>0$), here we consider the case $\nu=0$. In the case of a homogeneous system, Eq.~\eqref{eq:observable} is independent of the position $x$ and we may substitute $\hat{O}(x,t)$ by $\int \d x' \hat{O}(x')/L$. Using $\hat{O}(x')=\e^{-i\hat{P}x'}\hat{O}(0)\,\e^{i\hat{P}x'}$, since the state $\ket{\nu'}$ has a well-defined momentum $P_{\nu'}$, we obtain the selection rule $P_{\nu}=P_{\nu'}=0$. Hence, non-zero momentum excitations \textit{cannot} be probed (for instance, single quasi-particle excitations) by one-point functions. In contrast, for a disordered system, translation invariance is broken, which lifts this selection rule and \emph{any} excitation can be probed, including the low-lying single quasi-particle excitations, provided $\braket{0| \hat{O}(x)|\nu'} \neq 0$.

In the following, we are also interested in two-point functions such as the one-body correlation function $g_1(x,y,t) = \braket{\hat{a}^{\dagger}(x,t) \hat{a}(y,t)}$. In this case, the QSF reads as
\begin{equation}
\begin{split}
G(k,k';\omega) &= 2\pi\sum_{\nu,\nu',\mu} \rho_{\textrm{i}}^{\nu' \nu} \delta(E_{\nu'}-E_{\nu}-\omega)\int\d x\,\e^{-ikx}\\
&\times\braket{\nu | \hat{O}_1(x)| \mu} \int\d y\,\e^{-ik'y}\braket{\mu| \hat{O}_2(y)| \nu'}.
\end{split}
\end{equation}
For a translationally invariant system, it yields the selection rule $P_{\nu'} = P_{\nu}$ ($=0$ for weak quenches) similarly as above. 
The observed resonances of the quench spectral function are generated by operators creating single quasi-particle excitations with equal and opposite momenta, $\ket{\nu'=(\mu,\mu')}$ with $P_{\mu} = -P_{\mu'}=k$. It yields a frequency resonance at $E_{\nu'}=2E_k$, i.e. twice the energy of a single quasi-particle excitation. In disordered systems as considered here, this selection rule is again lifted and one expects to directly probe single-particle excitations at low energies. 

Compared to the clean case, besides allowing us to use one-point functions as mentioned above, the disorder introduces two main effects visible on the spectral features probed by the QSF. First, the energy resonances are randomly shifted from their clean-system counterparts for each disorder realization.
When we disorder average the QSF,
this eventually leads to a broadening in energy of the spectral lines, as compared with the corresponding clean system.
Second,
the disorder also induces a broadening of the spectral features in momentum space. 
This broadening is the result of a combination of scattering due to the disordered potential and localization in real space, which translates into delocalization in momentum space and therefore an additional broadening of the spectral features, see Appendix~\ref{app.pt} for details.  
Similar effects occur in the context of standard pump-probe spectroscopy techniques for spectral functions including the single-particle spectral function~\cite{knap2010Excitations} and the dynamical structure factor~\cite{roux2013Dynamic}.

As for any spectroscopic approach, the precise transitions probed will depend crucially upon the overlap $\rho_{\rm{i}}^{\nu'\nu} \bra{\nu} \hat{O}(x)\ket{\nu'}$ in Eq.~\eqref{eq:QSF_disorder_full_expression} being non-zero. This means that both the density matrix coherences, which contain information about the initial state, and the choice of the observable $\hat{O}$ play a key role in selecting the transitions which contribute to the QSF. In particular, different choices of observable allow us to probe different spectral features, see Sec.~\ref{sec.strong_int}.

\subsection{Numerical simulations}
\label{sec.results1}

In all of the following, we numerically simulate a disordered Bose-Hubbard chain of length $L=47$ with open boundary conditions using tensor network methods. We generate out-of-equilibrium dynamics following a global quench of the Hamiltonian from some initial $\hat{H}_{\rm{i}}$ to the post-quench $\hat{H}_{\rm{f}}$. We have tested several different quench protocols, including quenches of the hopping amplitude, on-site interaction strength and the disorder strength, as well as different quench amplitudes, and have observed qualitatively similar results in every case.
The initial state of the system is always the ground state of the Hamiltonian $\hat{H}_{\rm{i}}$, obtained using the density matrix renormalization group (DMRG) algorithm~\cite{schollwock2011Densitymatrix}, making use of techniques which help to prevent getting stuck in metastable states~\cite{White05,Hubig+15}. We truncate the local Hilbert space to a maximum of $N_{\rm{b}}=5$ bosons per site, which we have checked in detail is sufficient to yield well-converged results except at very small interaction strengths, which we do not consider in detail here.
The time-evolution of the observable $\hat{O}(x,t)$ is then computed with the post-quench Hamiltonian using the time-dependent variational principle~\cite{haegeman2016Unifying}, with the hybrid time evolution method~\cite{Goto+19,Paeckel+19,Chandra+20}.
We use a maximum bond dimension of $\chi=128$ and a maximum evolution time of $Jt_{\rm{max}}=20$, in line with the timescale accessible to many experimental platforms~\cite{neyenhuis2017Observation,garttner2017Measuring,kohlert2019Observation,zhou2020Quench,rispoli2019Quantum}.
In all data, we subtract the long-time average to remove peaks in the QSF which correspond to an irrelevant time-independent background signal. We apply a Hann window function to reduce boundary effects before taking the space-time Fourier transform. We then take the absolute magnitude of the QSF and average over disorder realizations in order to eliminate sample-to-sample variations, and finally we normalize the result. In Appendix~\ref{app.num}, we show some examples of the QSF with and without these processing steps to illustrate the effect of each. When combined with the relatively large system size used in this work, a modest number of disorder realizations proves to be more than sufficient, and demonstrates that our predictions are easily accessible to current generation experiments without requiring averaging over a number of disorder realizations that may be prohibitive in practice.

\section{Characterizing the quantum phases of the model from the QSF}
\label{sec.strong_int}
\begin{center}
\begin{figure*}[t]
\includegraphics[width=\linewidth]{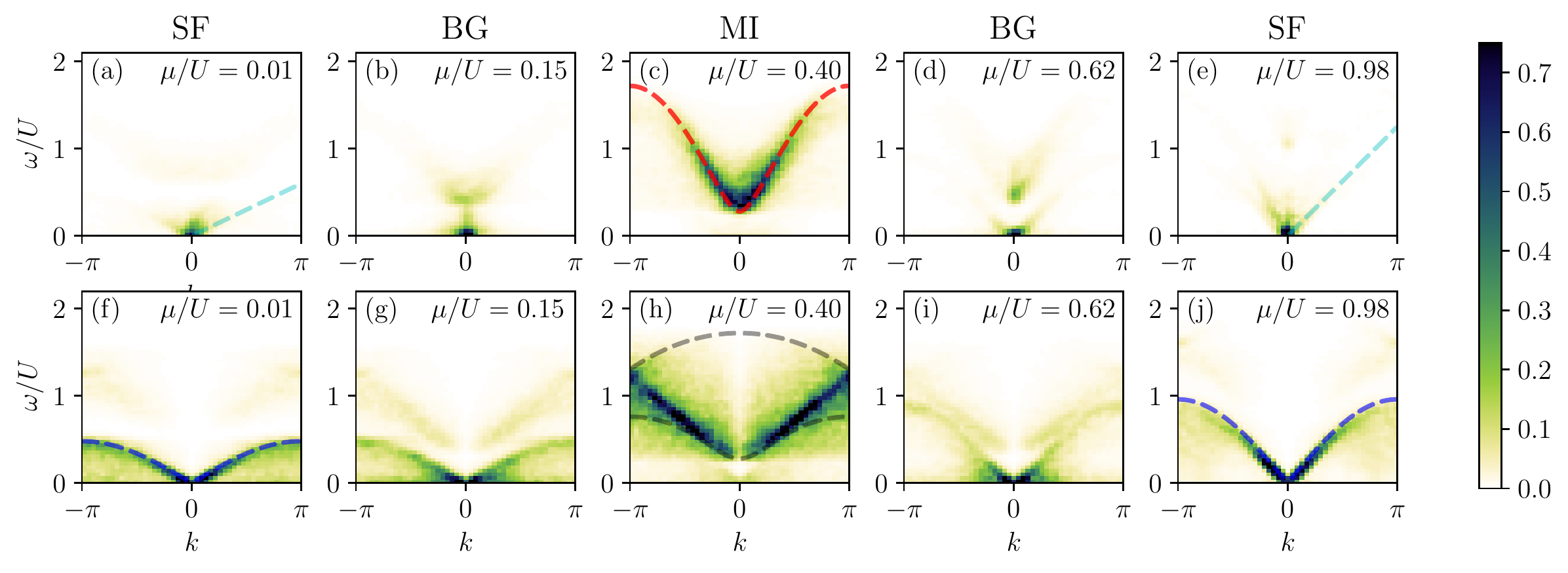}
\caption{
Disorder-averaged quench spectral functions (QSFs) of $g_1(x,t)$ (a-e) and $n(x,t)$ (f-j) with $U/J_{\rm{i}}=7.5$ and $\Delta/U=0.25$ after a quench from $J_{\rm{i}} = 1.0$ to $J_{\rm{f}}=0.9J_{\rm{i}}$, averaged over $N_{\rm{s}}=15$ disorder realizations and normalized. The superfluid is visible at $\mu/U=0.01$ and $\mu/U=0.98$, where the QSF of the one-body correlator shows a clear linear signal close to $k=0$ characteristic of a well-defined sound-like mode [cyan line in panels (a) and (e)], and can be described by an effective fermionic model which predicts the signal seen in the QSF of the density [blue line in panels (f) and (j)]. The Mott insulator phase is visible at $\mu/U=0.4$, and exhibits a clear gap in its spectrum, with an excitation band that closely matches the homogeneous Mott insulator dispersion relation [red line in panel (c)], while panel (h) shows that the QSF of $n(x,t)$ probes a continuum of excitations bounded by the sum of individual particle and hole dispersions (gray dashed line). Notably, both variables give a gapped response in the MI phase. The Bose glass, visible at $\mu/U=0.15$ and $\mu/U=0.62$, exhibits a coexistence of gapped and gapless excitations, with a broad continuum of excitations close to $\omega=0$ and weak gapped bands close to $\omega = U$.}
\label{fig.qsf_random}
\end{figure*}
\end{center}

In this section, we discuss the QSF of the one-body correlator $g_1(x,t)=\langle\hat{a}^{\dagger}(x,t) \hat{a}(0,t)\rangle$ and of the local density $n(x,t)=\langle\hat{n}(x,t)\rangle$. The former partially overlaps with and complements the discussion of Ref.~\cite{Paper1}. We work in the grand canonical ensemble and explore the regime where $J,\Delta\ll U$. In the strongly-interacting regime, all three phases (SF, BG and MI) can be explored by varying the particle number through the chemical potential, at fixed interaction strength $U$, hopping $J$, and disorder strength $\Delta$, see Fig.~\ref{fig.qsf_random}. 

In the following, we quench the hopping amplitude from $J_{\rm{i}}=1.0$ in the initial Hamiltonian to $J_{\rm{f}}=0.9J_{\rm{i}}$ in the final Hamiltonian. We work at interaction strength $U/J_{\rm{i}}=7.5$ and fixed (weak) disorder strength $\Delta/U=0.25$, allowing us to compare our numerical data to known analytical results in homogeneous systems \cite{cazalilla2003Onedimensional,cazalilla2004Differences,barmettler2012Propagation}. Importantly, the overall number of bosons is conserved during the quench.
Hence, the chemical potential changes with the Hamiltonian during the quench. Hereafter, the values of $\mu$ are relative to the post-quench Hamiltonian, see also Ref.~\cite{Paper1}.

The use of the one-body correlator is motivated by the fact that it is known to be a suitable probe of the excitation spectrum of both MI and SF phases in the clean system~\cite{villa2019Unraveling}, while the density was chosen as it is the simplest local probe which acts on only a single lattice site.
Both are standard observables, readily measured in experiments: The density can be measured directly from quantum gas microscopes with single-site resolution, while $g_1(x,t)$ is the Fourier transform of the momentum distribution as measured by time-of-flight imaging.
The resulting QSFs for these two variables are displayed in Fig.~\ref{fig.qsf_random} for a variety of values of the chemical potential $\mu/U$, which cover all three phases in both the high ($\overline{n} \geq 1$) and low ($\overline{n}<1$) filling regimes. We shall now discuss the behavior of the QSF in each phase separately.

\subsection{Superfluid}

\emph{One-body correlator} - The QSF of $g_1(x,t)$ in the SF phase is shown in Fig.~\ref{fig.qsf_random}(a) at low filling ($\overline{n}<1$) and Fig.~\ref{fig.qsf_random}(e) at higher filling ($\overline{n}>1$). In both cases, we find that the QSF is strongly peaked at $\omega/U=0$, indicating a gapless response. Additionally, for momenta close to $k=0$ we find that the QSF displays a well-defined linear slope, characteristic of phonon-like excitations which permit a global speed of sound. These observations are consistent with the expectation that the QSF of $g_1(x,t)$ should probe the excitation spectrum, as in translationally invariant systems~\footnote{By comparison with Ref.~\cite{villa2019Unraveling}, one may expect a signal at $2E(k)$ rather than the spectrum $E(k)$. As discussed above, however, due to the collective nature of the excitations (phonons) and the breaking of translation invariance due to the disorder, the QSF in fact displays a strong signature of the spectrum itself, at $E(k)$. There is an additional weak signal corresponding to $2E(k)$ visible in some data sets, originating from the same mechanism presented in Ref.~\cite{villa2019Unraveling}.}. 

The exact dispersion relation of the SF phase of the Bose-Hubbard model is not exactly known, even for the homogeneous system in the strongly interacting regime ($U/J\gg 1$) where multiple occupancy of individual lattice sites is strongly suppressed and the bosons become essentially hard core. 
However by restricting to the limit of low filling ($0 < \overline{n} < 1$), the local Hilbert space can be effectively truncated to two states ($n=0$ and $n=1$) and the homogeneous Bose-Hubbard chain can be mapped onto a model of spinless fermions~\cite{cazalilla2003Onedimensional,cazalilla2004Differences}.
Here, we consider SF regions with fillings both in the range $0 < \overline{n} < 1$, where we make use of the results of Refs.~\cite{cazalilla2003Onedimensional,cazalilla2004Differences}, and in the range $1 < \overline{n} <2$ where we apply the same logic. In the latter case, we truncate the local Hilbert space to two states (corresponding to single and double occupancy of a lattice site), which can be effectively described in terms of a spinless fermionic degree of freedom.

In the hard core limit, we obtain a fermionic tight-binding model whose elementary excitations consist of particle(p)-hole(h) pairs with individual momenta given by $\kp\in \phantom{.} ]-\pi;-\kf]\cup[\kf;\pi[$ and $\kh\in[-\kf;\kf]$, respectively, with the effective Fermi momentum $\kf = \bar{n}_{\rm{F}} \pi$, where $0 < \overline{n}_{\rm{F}} < 1$ is the average density of the spinless fermions. The excitation spectrum takes the form of a continuum given by $E(\kp;\kh)=-2\bef J(\cos \kp-\cos \kh)=4\bef J\sin(k/2)\sin(k/2+\kh)$ where we define $k=\kp-\kh$ and $\bef$ is the Bose enhancement factor. The latter takes the value $\bef=1$ for $0 < \overline{n} < 1$, and $\bef=2$ for $1 < \overline{n} < 2$. Close to the origin, such that $\kp\simeq \kh\simeq \kf$, the continuum reduces to a sound-like linear branch with velocity $2\bef J\sin(\kf)$.

For a two-point correlator such as $g_1(x,t)$, the main contribution to the QSF comes from the homogeneous term already present in the clean system~\cite{villa2019Unraveling} with an additional broadening of the observed signal due to the disorder, which we further discuss in Appendix~\ref{app.pt}. 
An additional strong contribution close to $k=0$ for frequencies $\omega>0$, caused by the sensitivity of the one-body correlator to the quasi-long-range order in the system, leads to a V-shaped continuum around $k=0$. Nonetheless, the lower edge of this continuum is linear close to $k=0$, with a gradient equal to the sound velocity~\cite{Paper1}, as indicated by the cyan line in Fig.~\ref{fig.qsf_random}. 

\emph{Density operator} - The QSF of the local density is shown in Fig.~\ref{fig.qsf_random}(f) at low filling ($\overline{n}<1$) and Fig.~\ref{fig.qsf_random}(j) at higher filling ($\overline{n}>1$). It exhibits many of the same features as the QSF of the one-body correlator, i.e.\ a gapless response and linear branches close to $k=0$. However, this signal has a different origin and is not due to the elementary excitation spectrum. 
The origin of this signal can instead be inferred from the following analytical argument which predicts the blue dashed lines seen in Fig.~\ref{fig.qsf_random}(f) and (j).
As shown in Ref.~\cite{villa2020Local}, a cosine-like dispersion of the form $-c\cos k$ (up to an irrelevant constant term, with $c$ an arbitrary prefactor which is independent of $k$)  would possess algebraic divergences at the energies given by $\omega=2c\sin(k/2)$, associated to the transition energies within the same excitation manifold. Following a similar argument for the excitation spectrum $E(k;\kh)=4\bef J\sin(k/2)\sin(k/2+\kh)$, relevant here, we find analogous divergences at $\omega=\pm 4\bef J\sin(k/2)$, see Appendix~\ref{app:calcul_4J_sink/2}. The latter are represented in Fig.~\ref{fig.qsf_random}(f) and (j) as the dashed blue lines for $\bef=1$ and $\bef=2$ respectively. The agreement with the QSF of the density is excellent.

\subsection{Mott insulator}

\emph{One-body correlator} - The QSF of the one-body correlator in the MI phase is shown in Fig.~\ref{fig.qsf_random}(c). It displays a clear qualitative difference from the result in the SF phase. Here, the QSF has essentially no weight at $\omega \approx 0$ and instead displays an excitation gap, above which there is a broad band of excitations. The latter closely match the excitation spectrum of a homogeneous MI (red dashed line), which may be obtained from the following argument.

In the hard core limit, excitations in the MI are excess particles and holes on top of the uniformly filled $\overline{n} \in \mathbb{Z}$ background. Here and in the following, we specify to an overall density $\overline{n}=1$ in the MI. For strong but finite interactions, $U/J \gg 1$, the excitations in the MI are essentially dressed particles and holes. The excitations can be formally understood in terms of Bogoliubov quasiparticles given by~\cite{barmettler2012Propagation}  
\begin{equation}
\label{eq:link_doublon_holon_elementary_excit_MI}
\begin{split}
\hat{\gamma}_{k,+}^{\dagger}&=u_{k}\,\hat{d}_{k}^{\dagger}+v_{k}\,\hat{h}_{-k}\\
\hat{\gamma}_{k,-}^{\dagger}&=u_{k}\,\hat{h}_{k}^{\dagger}-v_{k}\,\hat{d}_{-k},
\end{split}
\end{equation}
which are linear combinations of the doublons $(\hat{d})$ and holons $(\hat{h})$, and $u_{k}=O(1)$, $v_{k}=O(J/U)$. 
The doublons and holons are related to the physical bosons by 
\begin{equation}
\hat{a}_{j}^{\dagger}=\sqrt{2}Z_{j}\hat{d}_{j}^{\dagger}+Z_{j}\hat{h}_{j},
\end{equation}
where $Z_{j}=\prod_{j'<j}\exp\left(i\pi\sum_{\sigma=d,h}\hat{n}_{j',\sigma}\right)$ is the Jordan-Wigner string phase factor. 
Due to their fermionic statistics, there can only be a single doublon or holon on each lattice site. By diagonalizing the Hamiltonian in terms of the Bogoliubov quasiparticles, the dispersion relation above the threshold interaction strength $U/J > 4(\overline{n}+1)$ can be shown to be
\begin{equation}
E_{\pm}(k)=\mp J\cos k + \frac{1}{2}\sqrt{(U-6J\cos k)^2+32(J\sin k)^2}.
\label{eq.MI}
\end{equation}
In the MI, the signal seen in the QSF of the one-body correlator strongly resembles the excitation band seen in the clean system, given by $E_{+}(k) + E_{-}(-k)$ [red dashed line in Fig.~\ref{fig.qsf_random}(c)], with, additionally, the expected broadening due to the disorder. The energy gap at $k=0$ is very close to the value of the homogeneous MI, given by $E_{\textrm{gap}}/U = 1-6J/U$, consistently with the expected strong screening of weak disorder in long-wavelength limits~\cite{lugan2007b,lugan2011}.

\emph{Density operator} - By contrast, the QSF of the local density, shown in Fig.~\ref{fig.qsf_random}(h), displays a very different structure. It again exhibits a gapped response but, instead of the QSF being peaked around $k=0$, there are two almost linear branches within a weak continuum of excitations. This can be explained by expressing the density operator in terms of the Bogoliubov quasiparticles.
After restricting the Hilbert space to forbid a doublon and a holon from occupying the same site, the density operator may be written for unit filling as $\hat{n}_{j}=\hat{a}_{j}^{\dagger}\hat{a}_{j}=1+\hat{d}_{j}^{\dagger}\hat{d}_{j}-\hat{h}_{j}^{\dagger}\hat{h}_{j}$~\cite{barmettler2012Propagation}. 
The resulting expression contains a term which creates pairs of elementary excitations with total momentum $k$. These pairs of excitations lead to the continuum observed in Fig.~\ref{fig.qsf_random}(h). This continuum can be computed from Eq.~(\ref{eq.MI}), and its boundaries are indicated in Fig.~\ref{fig.qsf_random}(h) by dashed grey lines. The full continuum is in good agreement with the peaks in the QSF seen in the numerical data. 

\subsection{Bose glass}

\emph{One-body correlator} - The QSF of the one-body correlator $g_1(x,t)$ is shown in the BG phase in Fig.~\ref{fig.qsf_random}(b) (for filling $\overline{n}<1$) and Fig.~\ref{fig.qsf_random}(d) (for filling $\overline{n}>1$).
In both cases we find a strong gapless response close to the origin which originates from the locally SF regions, as well as a secondary gapped excitation band. This allows us to distinguish the BG from the MI through a non-zero value of the QSF close to the origin, $|G(0,0)|>0$. In addition, because the SF regions are localized in real space there are no long-wavelength ($k \approx 0$) phonons able to propagate throughout the lattice. This in turn means that there is no global speed of sound, and therefore that the QSF in the BG phase \emph{does not} display a linear soundlike branch close to $k=0$. It is instead characterized by an indistinct gapless feature close to the origin, which progressively becomes a sharp linear branch at the BG-SF transition. These contrasting properties enable us to clearly distinguish all three phases by their different responses.

\emph{Density operator} - The QSF of the local density is shown in the BG phase in Fig.~\ref{fig.qsf_random}(g) (for filling $\overline{n}<1$) and Fig.~\ref{fig.qsf_random}(i) (for filling $\overline{n}>1$). It exhibits a strong gapless response which resembles that seen in the clean superfluid, with an outer envelope given by the same expression $\omega=4\bef J\sin(k/2)$. However, there is significant broadening within this envelope, particularly close to $k=0$, indicative of strong localization in real space. There is also a gapped signal similar to that discussed above, which is, however, much weaker than the gapless response.

The BG has a complicated real-space structure. For instance, it may mix locally SF and locally MI regions. The highly inhomogeneous nature of the BG makes it difficult to speak about a single type of elementary excitation, and indeed one might expect the spectrum of the BG to be a superposition of the MI-like and SF-like spectra.
These features are, in addition, broadened by localization.
Deep in the BG phase, both the local MI and SF regions have finite sizes.
Since localization in real-space corresponds to delocalization in momentum space,
is results in a significant broadening of the signal.

\section{Quench spectroscopy at unit filling}
\label{sec.results2}

\begin{center}
\begin{figure*}[t!]
\includegraphics[width=\linewidth]{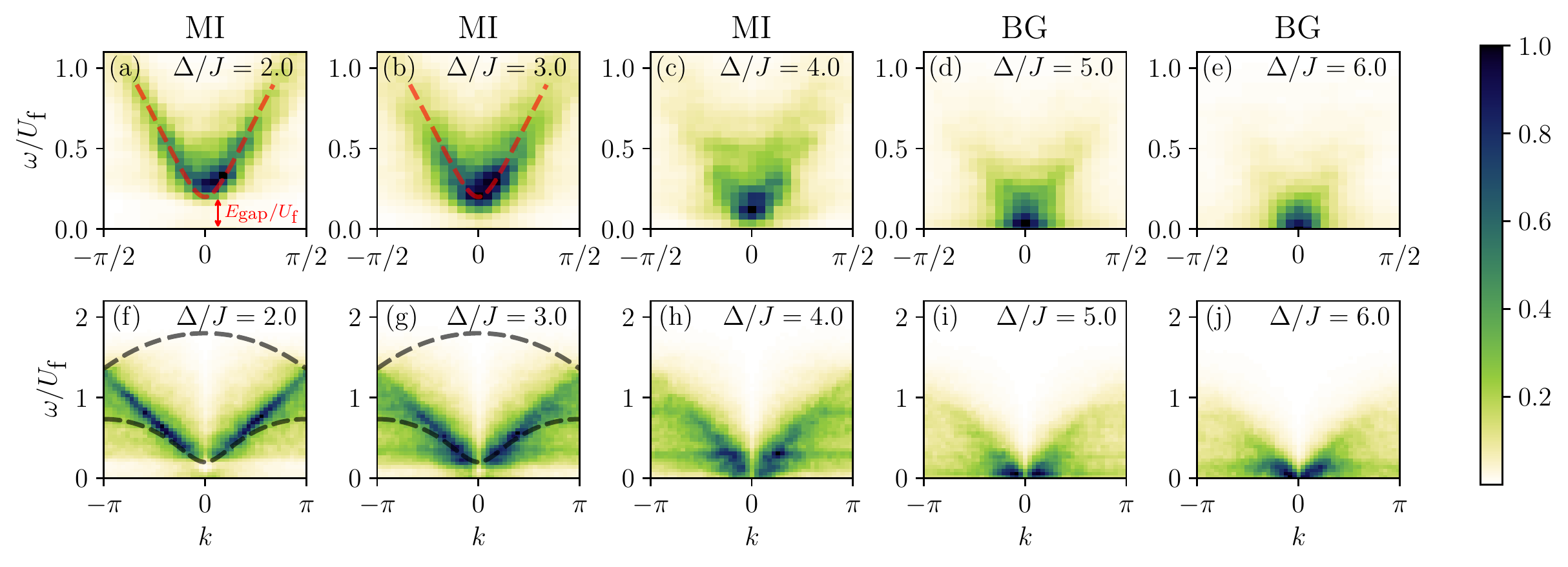}
\caption{The QSF in the disordered Bose-Hubbard model with fixed density $\bar{n}=1$ at $U_{\rm{f}}/J=7.5$ and a quench from $U_{\rm{i}}/J=0.9U_{\rm{f}}/J$, averaged over $N_{\rm{s}}=25$ disorder realizations. Panels (a-e) show the QSF of $g_1(x,t)$, while panels (f-j) show the QSF of $n(x,t)$. Note that we use different axis limits on each row to better highlight the features of interest. Here we start in the MI state for small values of $\Delta/J$, and as we increase the disorder the gap shrinks until we enter the BG phase at around $\Delta/J \approx 5$. The relatively sharp signals at small $\Delta/J$ values rapidly become broadened by the increasing disorder, and we observe a smooth transition from gapped to gapless behavior in both observables. The dashed red lines in panels (a) and (b) represent the dispersion relation of the homogeneous Mott insulator, while $E_{\textrm{gap}}$ labels the single-particle energy gap, and the dashed black lines in panels (f) and (g) represent the two-excitation continuum as seen previously in Fig.~\ref{fig.qsf_random}.}
\label{fig.qsf_rb2}
\end{figure*}
\end{center}

So far, we have varied the chemical potential $\mu$ and interaction strength $U$, while quenching the hopping amplitude $J$, and worked in the case where the leading effects of disorder could be captured by perturbation theory. We now go beyond this weak disorder limit, and turn to an investigation of the effects of varying the disorder strength $\Delta$ such that it becomes comparable to the on-site interaction $U$. We consider fixed commensurate filling $\overline{n}=1$, a situation commonly studied in numerical works~\cite{prokofev1998Comment,rapsch1999Density,roux2008Quasiperiodic,Goldsborough+15,Gerster+16,yao2016Superfluid}, and we demonstrate that in this regime quench spectroscopy also performs well in distinguishing all three phases, with the qualitative features in line with those discussed in Sec.~\ref{sec.strong_int}.

In all of the following, we again initialize the system in its zero-temperature ground state using DMRG, however this time we set $J=1$ and we consider quenches of the interaction strength to a final value of $U_{\rm{f}}/J$ from an initial value given by $U_{\rm{i}}/J=0.9U_{\rm{f}}/J$. Here we use $N_{\rm{s}}=25$ disorder realizations for all of the following. We show the results of several of quenches in Fig.~\ref{fig.qsf_rb2}. 

\subsection{Numerical Results}
\label{sec.rb_qsf}

\emph{One-body correlator} - The QSF of $g_1(x,t)$ (Fig.~\ref{fig.qsf_rb2}, top row) displays similar features to those shown in Sec.~\ref{sec.strong_int}, here at small $\Delta/J$ showing a well-defined signal which agrees well with the excitation band of a homogeneous MI (red dashed line). Again, as the disorder strength is increased, this band is broadened and its minimum moves to lower and lower frequencies, eventually becoming gapless. There is no transition into a superfluid phase in this strongly interacting, strongly disordered regime, and so the resulting QSF at large $\Delta/J$ values is largely featureless, characteristic of a strongly localized phase.

Contrary to the previous case of weak disorder where the Mott gap was only weakly modified by the random potential, here we can clearly see in Fig.~\ref{fig.qsf_rb2} the gradual closing of the Mott gap as the disorder strength is increased. This allows us to observe the Mott insulator to Bose glass transition via the closing of the gap, enabling the transition to be precisely located.
The QSF of $g_1(x,t)$ is also shown in Fig.~\ref{fig.Vs} for weaker interaction strengths, where there is a SF phase. The cyan lines indicate the linear fits to the soundlike modes used to extract the speed of sound in the SF phase. By contrast, panel (c) of Fig.~\ref{fig.Vs} is in the BG phase, where we do not see a well-defined linear branch emerge from $k=0$. Figure~\ref{fig.Vs}(d) shows the case of weak disorder for a value of $U/J$ that would be in the MI phase of the homogeneous system. The feature seen in this figure is not a linear soundlike branch, but is instead the MI excitation band which has been broadened by the disorder such that it becomes gapless.

\emph{Density operator} - As in Sec.~\ref{sec.strong_int}, the QSF of the density (Fig.~\ref{fig.qsf_rb2}, bottom row) again picks up a gapped signal within the two-excitation continuum. At small values of the disorder, this signal forms a sharp branch which lies inside the continuum boundary predicted in the absence of disorder (black dashed line). As the disorder strength $\Delta/J$ increases, this signal broadens in both momentum and frequency, leading to a broad continuum which eventually becomes gapless for large enough disorder strengths. At this point, the behavior of the system may again be understood from the fermionized tight-binding limit, and so exhibits the same gapless signal below an upper boundary given by $\omega=4\bef J\sin(k/2)$.

\subsection{Phase Diagram}
\label{sec.phase}

\begin{center}
\begin{figure}[t!]
\includegraphics[width=\linewidth]{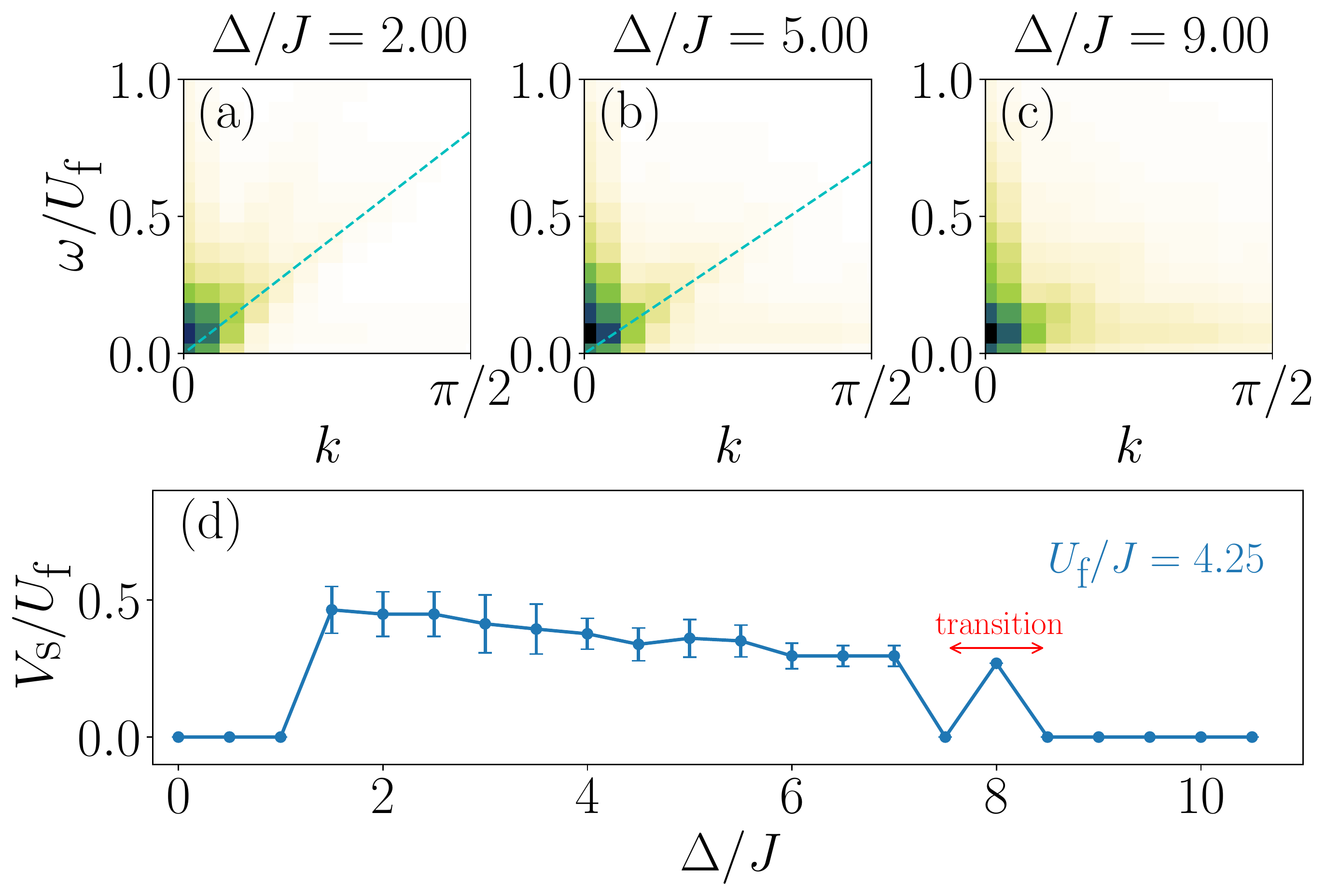}
\caption{(a-c) The disorder-averaged QSF of $g_1(x,t)$, shown for $U_{\textrm{f}}/J=4.25$ and a variety of values of $\Delta/J$ (indicated in the column label; the colour bar is the same as Fig.~\ref{fig.qsf_rb2}). The cyan lines represent linear fits close to the origin used to extract the sound velocity. Panels (a) and (b) are deep within the superfluid phase and the QSF exhibits a clear linear lower edge. Panel (c) is close to the SF-BG transition and exhibits significant broadening due to the disorder, though there is still a weak linear edge corresponding to a lower sound velocity than in panels (a) and (b). Panel (d) shows a plot of the extracted velocity $V_{\rm{s}}/U$ against $\Delta/J$ for $U_{\textrm{f}}/J=4.25$. The red line indicates the uncertainty in locating the SF-BG transition where $V_{\rm{s}}/U$ drops to zero: for disorder strengths $\Delta/J>U/J$, it can be very difficult to accurately fit the QSF with a linear slope, and fluctuations in $V_{\rm{s}}$ close to the transition are common.}
\label{fig.Vs}
\end{figure}
\end{center}

We can use the insights given in Sec.~\ref{sec.rb_qsf} to construct a phase diagram of the DBHM in the $(U/J,\Delta/J)$ plane at fixed hopping $J=1$. We have seen that the MI phase may be distinguished from both other phases via the existence of a finite excitation gap in the QSF of $g_1(x,t)$. This can be established in two ways: firstly by examining the zero-frequency amplitude of $|G(k=0,\omega=0)|$ as a function of $\Delta/J$ and $U_{\textrm{f}}/J$ (as used in Ref.~\cite{Paper1}), and secondly by directly extracting the excitation gap itself $E_{\textrm{gap}}(\Delta)$ (indicated in Fig.~\ref{fig.qsf_rb2}, and by the solid dark blue line in Fig.~\ref{fig.Vs}). 
When the gap is smaller than our numerical resolution, we instead compute the energy gap in the clean system ($\Delta/J=0$) and use the criterion $E_{\textrm{gap}}=\Delta$ to establish where in the phase diagram the Mott gap closes, as done in Ref.~\cite{prokofev1998Comment}. The result is indicated by the dashed dark blue line in Fig.~\ref{fig.Vs}. 

We can further use the existence of a finite speed of sound to distinguish the SF phase from both other phases. In the SF, the QSF of $g_1(x,t)$ is characterized by a well-defined linear slope close to $k=0$, the gradient of which is equal to the sound velocity in the system. By performing a linear fit to the QSF close to the origin, we can extract the sound velocity $V_{\rm{s}}$. The border between the SF and BG phases is given by the point at which $V_{\rm{s}}$ drops to zero, as in the BG there are no long wavelength phonons able to propagate throughout the entire lattice and consequently no speed of sound, see Ref.~\cite{Paper1}.

\begin{center}
\begin{figure}[t!]
\includegraphics[width=\linewidth]{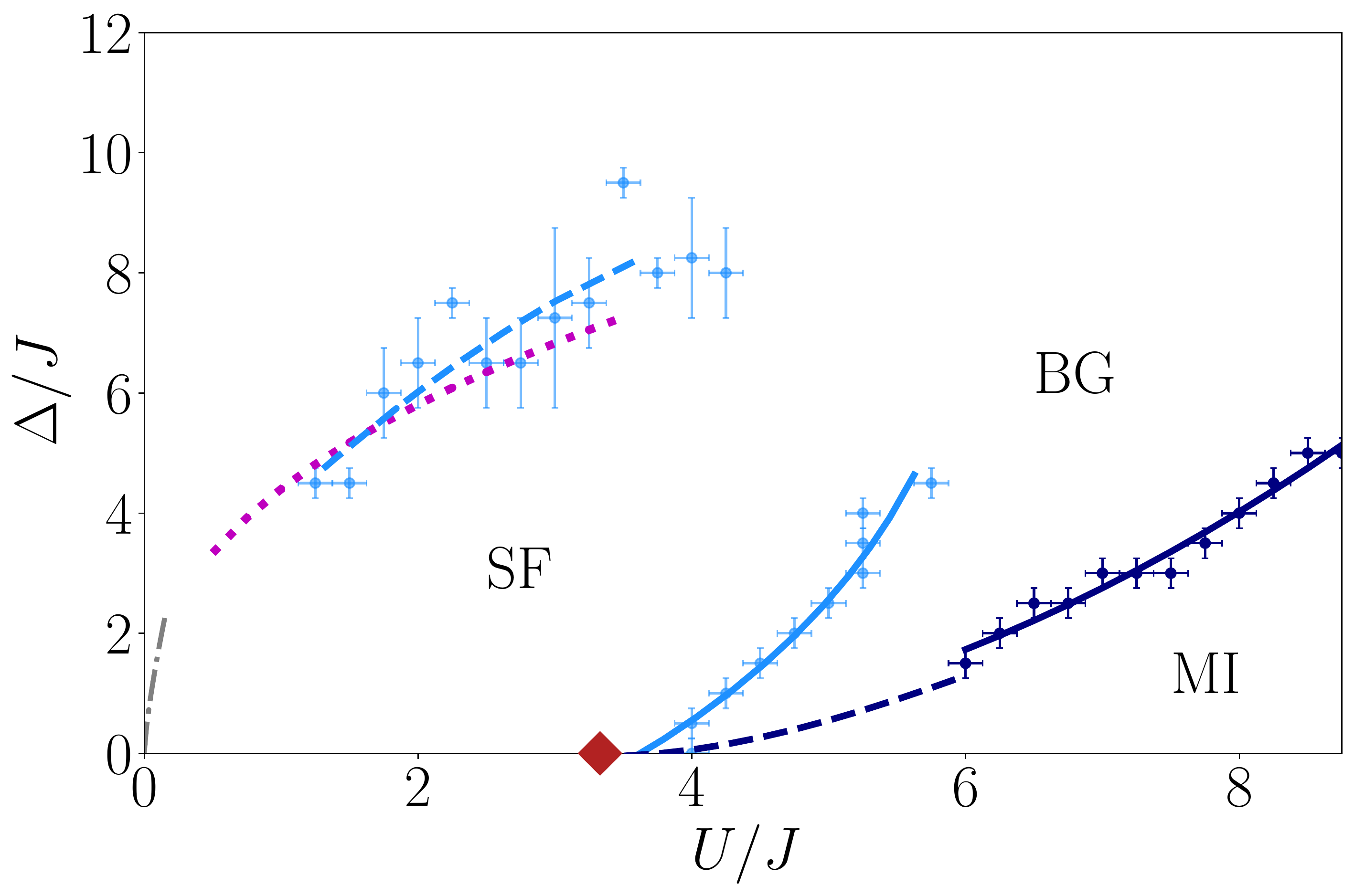}
\caption{Phase boundaries obtained from the QSF of the one-body correlator $g_1(x,t)$. The light blue line indicates the SF-BG boundary obtained from the speed of sound analysis, while the dark blue line shows the MI-BG boundary obtained from $E_{\textrm{gap}}\approx 0$. The red diamond indicates the MI-SF transition in the clean system ($\Delta/J=0$) at $U/J\simeq 3.3$, and the black dotted line is the estimated MI-BG transition based on the gap closing condition $E_{\textrm{gap}}=\Delta$, where the energy gap is measured in the clean system. The purple dotted line is an analytic expression for the upper boundary of the SF region from Ref.~\cite{Gerster+16}.
}
\label{fig.phase_Vs}
\end{figure}
\end{center}

The results of both of these measures are shown in Fig.~\ref{fig.phase_Vs}, where we demonstrate that through the QSF of $g_1(x,t)$.
The SF-MI transition in the clean case ($\Delta/J=0$) is denoted by the red diamond at $U/J \approx 3.3$~\cite{kashurnikov1996Mott,kuhner2000Onedimensional,ejima2011Dynamic}. As the Mott gap closes exponentially with decreasing $U/J$, we find that below approximately $U/J \approx 4$ the Mott gap is smaller than our frequency resolution, and so we cannot resolve it here. For large disorder strengths, the phase boundary must be taken as qualitative only, due to significant difficulties in performing the linear fits required to extract $V_{\rm{s}}$ caused by the disorder-induced broadening of the spectral function. Moreover, the results are significantly affected by finite size effects close to the SF-BG transition. In this parameter regime the typical size of the locally SF regions in the BG phase is close to the system size, and so even in the BG the spectrum strongly resembles that of the SF. Larger system sizes are required in this parameter regime to more clearly distinguish SF and BG phases through their long-wavelength behavior.

At small disorder, however, we find that the SF-BG phase boundary accurately matches that of other probes, in particular the one found using local quench spectroscopy as discussed below, see Fig.~\ref{fig.phase}.

\section{Local Spectral Function}
\label{sec.local}

The numerical results of Sec.~\ref{sec.strong_int} and~\ref{sec.results2} have shown that the QSF of the local density $n(x,t)$ is gapless in the BG and SF phases, and gapped in the MI. We now define an alternative spectral function based on the local density, this time a \emph{local spectral function} (LSF), which is the Fourier transform of Eq.~\eqref{eq:observable} in the time/frequency domain only 
\begin{align}
G(x,\omega) =2\pi\sum_{\nu,\nu'}\rho_{\rm{i}}^{\nu'\nu} \delta(E_{\nu'} - E_{\nu} - \omega) \braket{\nu | \hat{O}(x) | \nu'}.
\end{align}
This function retains spatially-resolved information about the excitation spectrum, and in particular is able to tell us if an individual lattice site hosts gapless excitations, enabling us to take a single disorder realization and establish which regions of the sample are gapless (locally SF regions) and which are gapped (locally MI regions). 

\begin{center}
\begin{figure}[t!]
\includegraphics[width=\linewidth]{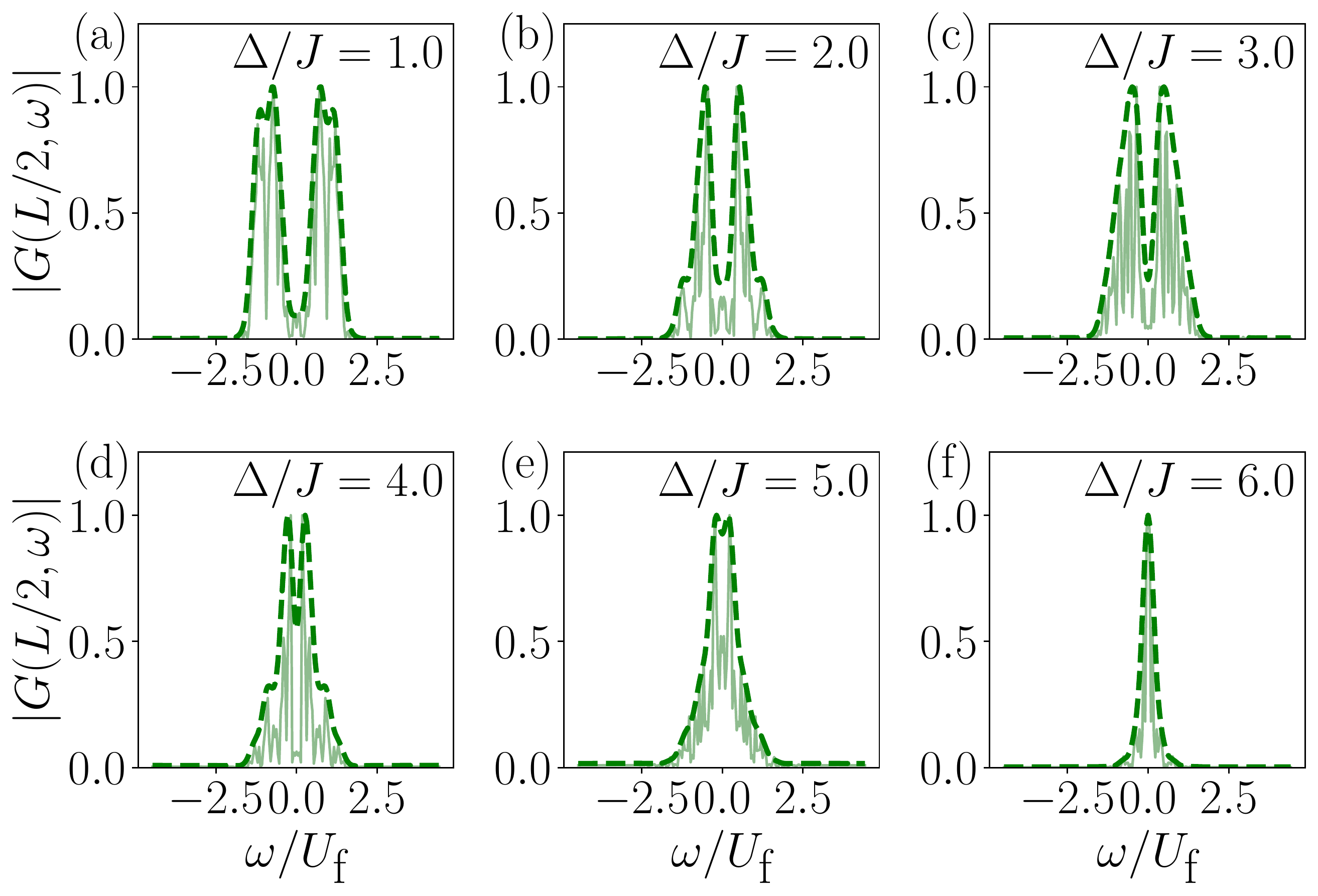}
\caption{The local spectral function $G(x,\omega)$ with $x=L/2$ shown for $U_{\rm{f}}/J=7.0$ and disorder strengths from $\Delta/J=1.0$ to $\Delta/J=6.0$. Each plot is a typical result from a single disorder realization. The solid green line is the data, and the dashed green line is a Gaussian convolution, necessary to smooth out the random effects of disorder. As the disorder strength is increased, the peaks of the LSF move towards $\omega=0$.
}
\label{fig.threshold}
\end{figure}
\end{center}

In Fig.~\ref{fig.threshold} we show several examples of the normalized LSF obtained on the central site of the chain at $U_{\textrm{f}}/J=7.0$ for four different disorder strengths across the MI-BG transition. The signal is rather noisy due to the amplitude of the peaks being disorder dependent, so we apply a Gaussian convolution (dashed lines) to smooth the signal before performing any further analysis. As $\Delta/J$ increases, the peaks in the LSF gradually move closer to zero until the Mott gap eventually closes. Note that the LSF at a single lattice site cannot by itself uniquely identify the phase, as in the BG phase, different lattice sites within a single disorder realization may exhibit gapped or gapless excitations. To identify each phase, we must analyze the behaviour of the LSF across all lattice sites.

A lattice site is defined as hosting gapless excitations if the $\omega=0$ peak is above some threshold value such that $|G(x,\omega=0)|> \varepsilon$, and gapped otherwise. It is necessary to impose this condition because even lattice sites in the MI may display a small signal at $\omega=0$. This is due in part to our use of a broad Gaussian convolution to smooth the signal, which may contribute a spurious zero frequency response, and partly due to the normalization used. The amplitude of the peaks of the LSF is disorder dependent, and so even in the MI there exist rare disorder realizations where the amplitude of the peaks is extremely small. Normalizing the LSF has the effect of amplifying noise in the signal, which can contribute a small zero-frequency `peak' and must be filtered out by the choice of an appropriate threshold. We pick the threshold $\varepsilon = 0.7$, which we empirically find to be the lowest value which correctly reproduces $\xi/L=1$ in the homogeneous Mott insulator phase ($\Delta/J=0$) for all points where the single-particle energy gap is within our numerical resolution. Small changes of $\varepsilon$ do not qualitatively change the phase boundaries.

By establishing which lattice sites are host to gapless and gapped excitations, we can extract a lengthscale $\xi$ defined by the typical size of the locally MI regions, or equivalently, the typical distance between the locally SF regions. 
In the MI, we have $\xi/L=1$ (i.e. the `MI regions' are the size of the entire system), while in the SF we have $\xi/L\approx 0$. In the BG, on the other hand, due to the presence of both types of local order, we expect $0 < \xi/L < 1$. 
As with any other measure, finite-size effects play a role here: In the BG phase but close to the MI-BG transition where the typical size of the MI regions may be larger than the size of the lattice simulated, this method will return the MI result of $\xi/L \approx 1$. Likewise, close to the SF-BG transition where the typical size of the SF regions are larger than the lattice size, this measure will return the SF result of $\xi/L \approx 0$. Consequently, we expect to slightly overestimate the SF and MI regions in the phase diagram as compared with the thermodynamic limit, a shortcoming shared by all simulations on finite-size systems.

\begin{center}
\begin{figure}[t!]
\includegraphics[width=\linewidth]{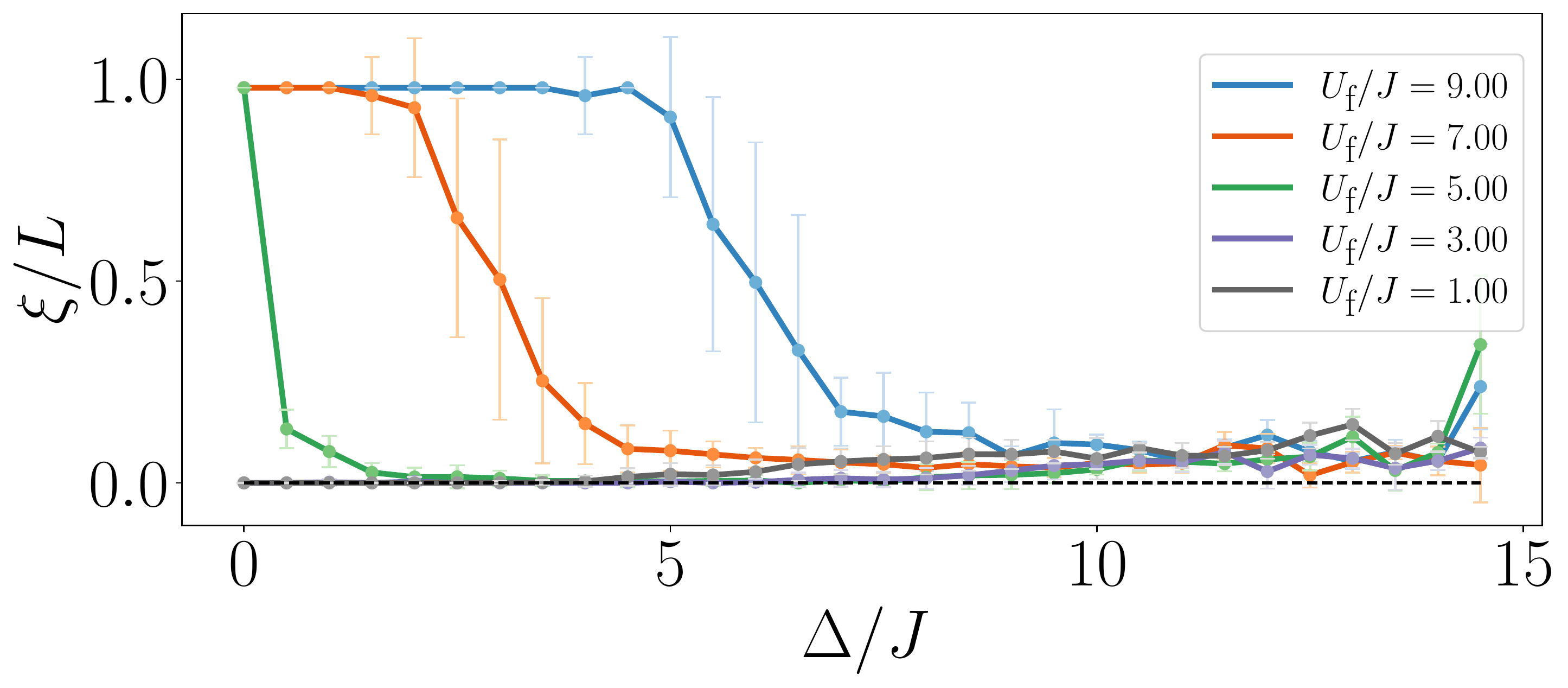}
\caption{The SF region spacing $\xi/L$ plotted against disorder strength $\Delta/J$ for five different values of final interaction strength $U_{\textrm{f}}/J$, averaged over $N_{\rm{s}}=25$ disorder realizations. In the MI, $\xi/L = 1$, while in the SF $\xi/L = 0$. In the BG phase, $0 < \xi/L < 1$. The error bars show the standard deviation of $\xi/L$ across disorder realizations: They are close to zero in the MI and SF phases, but abruptly increase upon entering the BG phase. This illustrates that the variance of the distribution of gapped/gapless regions, as well as the average, acts as an order parameter for the BG phase. This is the same data used to construct Fig.~\ref{fig.xi2}. 
}
\label{fig.xi}
\end{figure}
\end{center}

After disorder averaging, the LSF of the density operator is able to extract the typical spacing of gapless regions. To identify each phase, we consider quenches to a final interaction strength $U_{\rm{f}}$ from an initial interaction strength $U_{\rm{i}} = 0.9 U_{\rm{f}}$. Our results for $\xi/L$ are shown in Fig.~\ref{fig.xi} for several different values of $U_{\textrm{f}}/J$. Here we average over $N_{\rm{s}}=25$ disorder realizations. The error bars indicate the standard deviation over disorder realizations. We can see clearly that $\xi/L=1$ in the MI phase and $\xi/L \approx 0$ in the SF phase, as expected, while this quantity takes on intermediate values in the range $0 < \xi/L < 1$ in the BG phase, with large error bars reflecting the changes in the distribution of SF regions from sample to sample. The curves with small interaction values ($U_{\textrm{f}}/J=1.0$ and $3.0$) have $\xi/L=0$ for small disorder values in the SF phase, and become non-zero as the disorder strength is increased and the system undergoes a phase transition into the BG. By contrast, the curves for larger interaction values start in the MI phase with $\xi/L=1$ for $\Delta/J=0$ and undergo transitions into the BG phase, where $\xi/L$ becomes smaller than one. 

\begin{center}
\begin{figure}[t!]
\includegraphics[width=\linewidth]{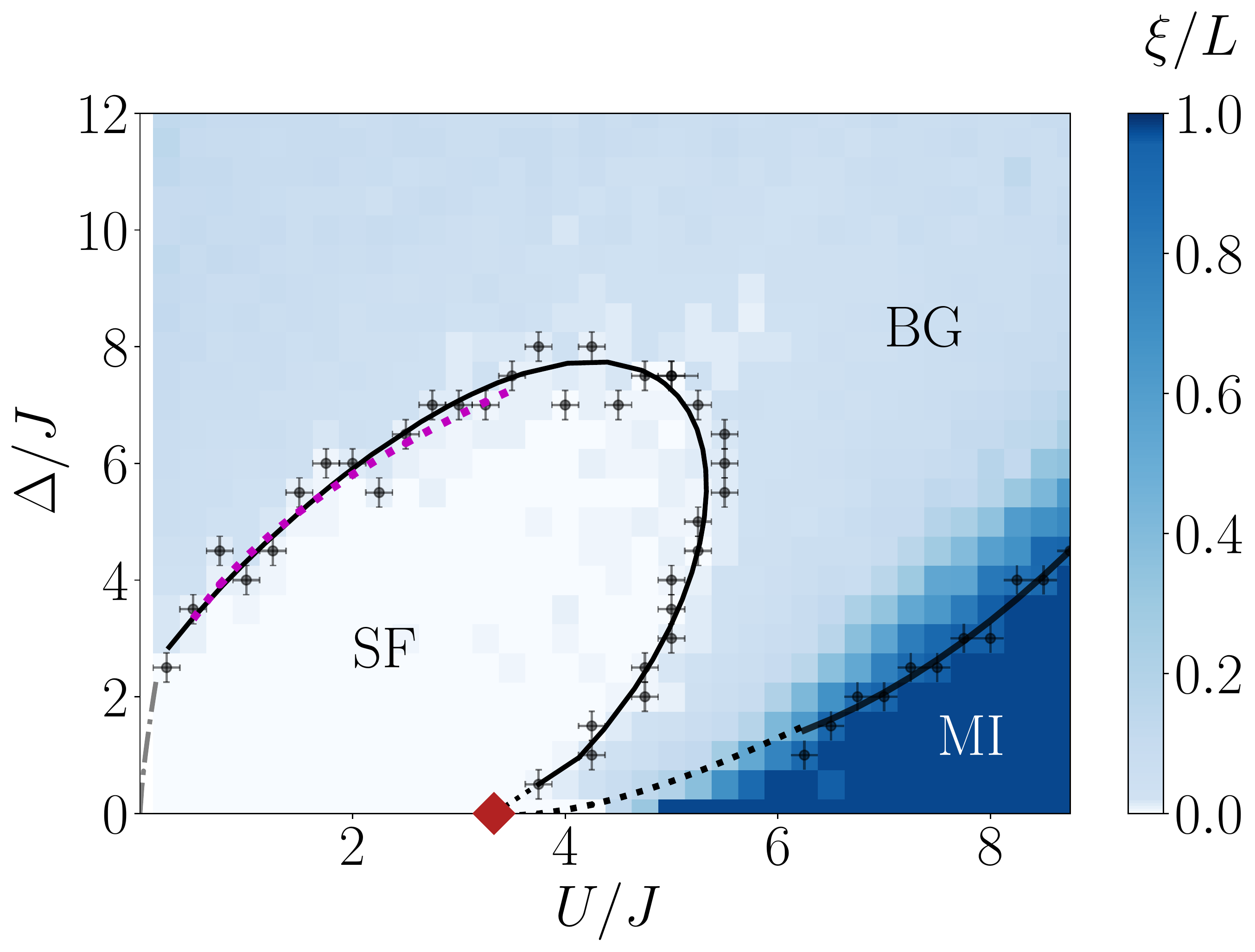}
\caption{The phase diagram in terms of the SF region spacing $\xi/L$, obtained from the LSF of the local density for system size $L=47$ and averaged over $N_{\rm{s}}=10$ disorder realizations. The SF phase is identified from the condition $\xi/L \approx 0$, the MI from $\xi/L = 1$ and the BG by $0 < \xi/L < 1$. The black lines are fits to the data points intended as a guide to the eye. Due to the exponential closing of the Mott gap close to the critical point of the clean system (red diamond) and our finite numerical resolution, below a certain point we are unable to resolve the Mott gap: Here we instead indicate the MI-BG boundary using the condition $E_{\textrm{gap}}(\Delta=0)=\Delta$ (black dotted line). The purple dotted line is an analytic expression for the upper boundary of the SF region from Ref.~\cite{Gerster+16}.
}
\label{fig.xi2}
\end{figure}
\end{center}

By repeating this procedure we are able to reconstruct the entire phase diagram using the SF region spacing $\xi/L$ as our order parameter, as shown in Fig.~\ref{fig.xi2}. As compared with the sound velocity discussed in Sec.~\ref{sec.results2}, the parameter $\xi/L$ is easier to numerically compute and gives much more robust results across the entire phase diagram.
To obtain the SF-BG boundary shown on the phase diagram, we scanned along horizontal (for the lower boundary) and vertical (for the upper boundary) lines and extracted the first point at which $\xi>1$, i.e. the point at which the typical size of locally gapped regions exceeds the lattice spacing. The error bars are given by the resolution of the underlying grid, except in rare cases where $\xi/L$ displays weakly non-monotonic fluctuations in the vicinity of the boundary, in which case we average over several points as indicated by the error bar. The black line is a guide-to-the-eye parametric fit of these points. The MI-BG boundary is similarly found by scanning down vertical lines in the phase diagram and extracting the first point at which $\xi \geq L-1$, where the typical size of a SF region becomes of the order of a single lattice site.

We have verified that both the SF-BG boundary and the MI-BG boundary shown in Fig.~\ref{fig.xi2} are in excellent agreement with existing numerical work using other methods~\cite{prokofev1998Comment,Gerster+16,yao2016Superfluid}.
As the system size reached in those works is larger than ours, by comparison we slightly overestimate the size of the ordered phases.
Note also that the MI-BG boundary does not precisely agree close to $U/J \sim4$ due to the Mott gap becoming smaller than our numerical resolution, however we see good agreement at larger interaction strengths. We find a remarkably accurate quantitative agreement between the upper boundary of the SF region and the analytical formula obtained in Ref.~\cite{Gerster+16} from a fit of their numerical data (purple line in Fig.~\ref{fig.xi2}). 

\section{Discussion/Conclusion}
\label{sec.conclusion}
In this work, we have extended the quench spectroscopy technique previously used to study the excitation spectra of homogeneous systems to the case of disordered systems. By studying the dynamics following a global quench, we have distinguished all three zero-temperature phases of the disordered Bose-Hubbard chain. We focused on two observables readily observable in current generation experiments, the density and the one-body correlator, and we have shown that spectral properties (such as the elementary excitation spectrum) and thermodynamic properties (such as the sound velocity) of the various phases can be obtained. 
Quench spectroscopic methods allow for direct measurement of the speed of sound in ultracold atomic gas experiments, which we have demonstrated in a theoretically challenging regime. We have also introduced the \emph{local} spectral function,
which paves the way for studies of the SF-BG transition by studying the growth of superfluid regions, particularly in dimensions greater than one, where the non-equilibrium dynamics are extremely challenging to simulate using exact numerical methods. The measurements proposed here are straightforward and may be conveniently implemented in a wide variety of systems beyond the Bose-Hubbard model, in experimental platforms including spin chains~\cite{smith2016Manybody}, fermionic systems~\cite{Schreiber+15} as well as continuous models where the BG phase has recently been shown to exist~\cite{yao2020LiebLiniger,sbroscia2020,gautier2021Strongly}. This highlights the potential for a more widespread adoption than other spectroscopic techniques such as momentum-resolved Bragg spectroscopy which require more finely-tuned experimental setups.

One promising avenue for future work is the extension of quench spectroscopy to quasiperiodic systems, where instead of generating the disordered potential in a random manner, is it instead generated using bichromatic potentials with incommensurate wavelengths~\cite{damski2003Atomic,Fallani+07,guarrera2007Inhomogeneous}. Quasiperiodic systems have received a great deal of study in both experimental and theoretical~\cite{roux2008Quasiperiodic,roux2013Dynamic} contexts, and are extremely interesting from a theoretical point of view due to their lack of Griffiths rare-region effects. This technique may also have applications towards the study of spectral features in many-body localized systems~\cite{Nandkishore+14}.

\acknowledgments
Numerical calculations were performed using HPC resources from CPHT
and HPC resources from GENCI-CINES (Grants 2019-A0070510300 and 2020-A0090510300).
We acknowledge use of the QuSpin~\cite{weinberg2017quspin,weinberg2019quspin} and TenPy~\cite{Hauschild+18} packages.

\appendix

\section{Weak quenches and the role of temperature in standard spectroscopy}
\label{app.temp}

Here, we discuss the role played by the temperature in pump-probe spectroscopy and compare it to the strength of the quench performed in quench spectroscopy. The following discussion also applies to clean systems. We recall that pump-probe spectroscopy techniques rely on the assumption that the system is weakly perturbed by a pump, such that by using linear response theory, its out-of-equilibrium properties can be related to the correlation functions at equilibrium of the unperturbed system, via the Kubo formula. Spectral properties are probed through spectral functions defined in a Gibbs thermal ensemble, and which are given by space-time Fourier transforms of unequal time correlators. One commonly used spectral function is the so-called dynamical structure factor, defined for a particle model initially at equilibrium with a thermal bath of finite temperature $\beta^{-1}$ as
\begin{equation}
\label{eq:DSF_finite_T}
S(k,\omega)=2\pi\sum_{n,m}\frac{\e^{-\beta E_{n}}}{Z}\abs{\bra{m}\hat{n}_{k}\ket{n}}^{2}\,\delta(\omega-E_{m}+E_{n}),
\end{equation}
with $Z=\sum_{n}\e^{-\beta E_{n}}$. At zero temperature it reduces to
\begin{equation}
\label{eq:DSF_T=0}
S(k,\omega)=2\pi\sum_{m}\abs{\bra{m}\hat{n}_{k}\ket{0}}^{2}\,\delta(\omega-E_{m}),
\end{equation}
where the ground state energy has been chosen such that $E_{0}=0$. 
Importantly, at zero temperature only the transitions to the ground state are probed by the dynamical structure factor. In contrast, in quench spectroscopy the initial state is out-of-equilibrium (with respect to the post-quench Hamiltonian), therefore the off-diagonal elements of the initial density matrix in the eigenbasis of the post-quench Hamiltonian are non-zero, and all transitions between excited states $E_{n}-E_{m}$ can be probed even at zero temperature. As we start from the ground state of the pre-quench Hamiltonian and perform a weak global quench, the post-quench state (initial state for the dynamics) is close to the pre-quench state, and only the low-energy excited states on top of it are targeted. This is to be compared to the situation in standard spectroscopy close to equilibrium at finite but small temperature, where only the low-energy states contribute significantly to the spectral functions. The effect of temperature in standard pump-probe spectroscopy hence plays a similar role as the strength of the quench in quench spectroscopy, where stronger quenches excite more and more energy levels and allow additional transition lines to appear in the QSF. Note however that a pure state is generated in QSF in contrast to finite temperatures, which generate statistical mixed states.

\section{Perturbation theory}
\label{app.pt}
In this Appendix we analytically explore the roles played by weak disorder on the spectral features probed by the QSF. For simplicity of the notations we restrict to one dimension, and we denote by $\left\{\ket{\nu}\right\}$ (Greek indices) the disorder-dependent eigenstates of the post-quench Hamiltonian and $\left\{\ket{n}\right\}$ (Latin indices) the eigenstates of the corresponding clean system ($\Delta=0$). 
We consider the case where our observable is a one-point function $\hat{O}(x,t)$.
Note that since translation invariance is broken by the disorder, we have $[\hat{H},\hat{P}]\neq 0$, therefore the energy eigenstates are no longer momentum eigenstates. For a single disorder realization, the dynamics are given by 
\begin{equation}
G(x;t)
=\text{Tr}[\e^{-i\hat{H}t}\,\hat{\rhoi}\,\e^{i\hat{H}t}\,\e^{-i\hat{P}x}\,\hat{O}\,\e^{i\hat{P}x}],
\end{equation}
where for simplicity we write $\hat{O}=\hat{O}(0,0)$. Decomposing onto the energy eigenbasis, and inserting a completeness relation it can be written as
\begin{equation}
\label{eq:general_QSF_without_simplification_with_disorder}
\begin{split}
G(x;t)&=\sum_{\nu,\nu'}\rhoi^{\nu'\nu}\,\e^{i(E_{\nu}-E_{\nu'})t}\,\bra{\nu}\e^{-i\hat{P}x}\,\hat{O}\,\e^{i\hat{P}x}\ket{\nu'}.
\end{split}
\end{equation}
Importantly, for a clean system the Hamiltonian and the momentum operator can be diagonalized simultaneously such that 
\begin{equation}
\bra{n}\e^{-i\hat{P}x}\hat{O}\,\e^{i\hat{P}x}\ket{n'}=\e^{i(P_{n'}-P_{n})x}\bra{n}\hat{O}\ket{n'}.
\end{equation}
However, this is no longer true for the disordered system in terms of the disorder-dependent eigenstates, and Eq.~\eqref{eq:general_QSF_without_simplification_with_disorder} cannot be simplified further on general grounds. Consequently, after taking the space-time Fourier transform we no longer obtain the selection rule in momentum that linked the frequency resonances to the momentum ones and yielded the sharp spectral features in the clean system~\cite{villa2019Unraveling}. When translation invariance is broken by the disorder, the QSF for a single disorder realization reads as
\begin{equation}
\begin{split}
G(k;\omega)&=2\pi\sum_{\nu,\nu'}\rhoi^{\nu'\nu}\,\delta(E_{\nu'}-E_{\nu}-\omega)\\
&\quad\times\int\d x\,\e^{-ikx}\bra{\nu}\e^{-i\hat{P}x}\,\hat{O}\,\e^{i\hat{P}x}\ket{\nu'}.
\end{split}
\end{equation}
To gain further insight on the role played by the disorder, let us restrict to weak disorder so as to assume that we can treat the disorder term as a small perturbation. To first order in perturbation theory, the disorder-dependent eigenstates and energies are related to their clean counterpart by
$\ket{\nu}=\ket{n}+\sum_{m\neq n}\mc{D}_{nm}^{\star}\ket{m}$ and $E_{\nu}=E_{n}+V_{n}$, where we define the disorder-dependent quantities
$\mc{D}_{nm}:=\frac{\bra{n}\hat{H}_{\rm{pert}}\ket{m}}{E_n-E_m}=-\mc{D}^{\star}_{mn}$ and $V_{n}=\bra{n}\hat{H}_{\rm{pert}}\ket{n}$. We also decompose $\hat{\rhoi}=\hat{\rho}_{\rm{i},(0)}+\delta\hat{\rhoi}$ where $\hat{\rho}_{\rm{i},(0)}$ refers to the homogeneous system in the absence of disorder. Using that the energy eigenstates of the homogeneous system are also eigenstates of the momentum operator, we obtain at first order
\begin{widetext}
\begin{equation}
\label{eq:QSF_to_comment_physical_effects}
\begin{split}
G(k;\omega)&\simeq (2\pi)^{2}\sum_{n',n}\delta\left(E_{n'}-E_{n}-\omega+V_{n'}-V_{n}\right)\\
&\quad\times\left[\delta(P_{n'}-P_{n}-k)\bra{n}\hat{O}\ket{n'}\left(\delta\rhoi^{n'n}+\sum_{m\neq n'}\rho^{mn}_{\rm{i},(0)}\mc{D}_{n'm}+\sum_{m\neq n}\rho^{n'm}_{\rm{i},(0)}\mc{D}_{nm}^{\star}\right)\right.\\
&\left.\qquad+\rho^{n'n}_{\rm{i},(0)}\left(\sum_{m\neq n'}\delta(P_m-P_n-k)\bra{n}\hat{O}\ket{m}\mc{D}_{n'm}^{\star}+\sum_{m\neq n}\delta(P_{n'}-P_m-k)\bra{m}\hat{O}\ket{n'}\mc{D}_{mn}\right)\right].
\end{split}
\end{equation}
\end{widetext}
Let us comment on this result. First, by breaking translation invariance the disorder prevents the selection rule $P_{n'}=P_{n}$ that appears in the homogeneous system, and therefore enables the use of one-point functions to probe spectral properties at non-zero momentum. This is why the zeroth-order term in Eq.~\eqref{eq:QSF_to_comment_physical_effects}, corresponding to the clean system, vanishes.  Second, the disorder shifts the resonances of the energy selection rule. For each disorder realization, the term $V_{n'}-V_{n}$ varies and after disorder averaging this leads to a broadening of the spectral features in the frequency domain. We also note that each branch is weighted by the disorder and therefore the amplitude of the QSF varies significantly from sample to sample. This is explicitly visible in Fig.~\ref{fig.num}, where we only recover smooth spectral features after disorder averaging. Last but not least, Eq.~\eqref{eq:QSF_to_comment_physical_effects} also highlights that the disorder modifies the standard momentum selection rule $\delta(P_{n'}-P_{n}-k)$ by replacing either $P_{n'}$ or $P_{n}$ by an arbitrary (but distinct) momentum $P_m$ [see terms in the last line]. This leads to an additional broadening of the spectral features in momentum space.

\section{Numerical Considerations}
\label{app.num}

\begin{center}
\begin{figure}[t!]
\includegraphics[width=\linewidth]{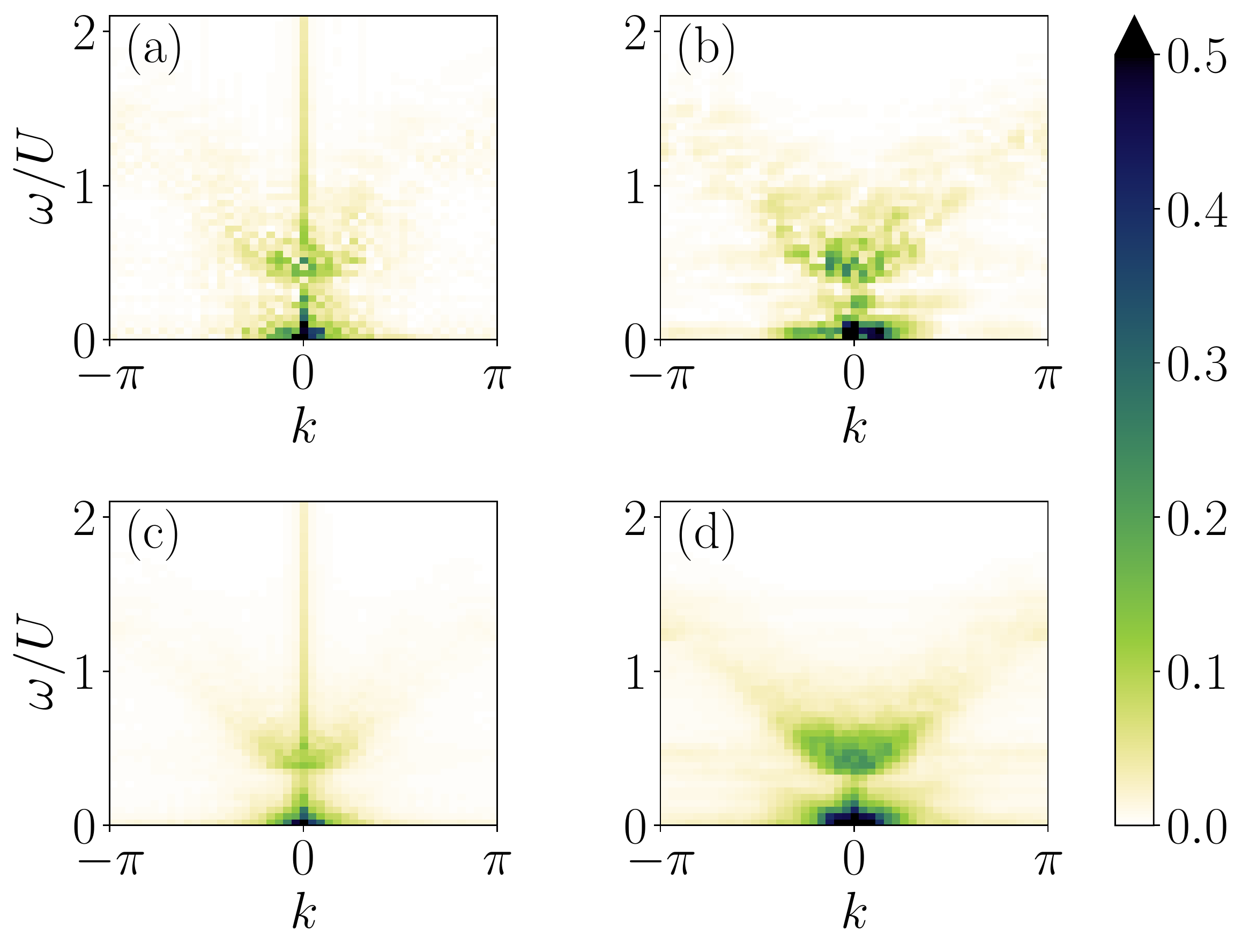}
\caption{The effects of different processing steps on the appearance of the final QSF. All plots show the QSF of $g_1(x,t)$ with $U/J_{\textrm{i}}=7.5$ and $\mu/U=0.15$. Panels (a) and (b) show the QSF of a single disorder realisation, without (a) and with (b) the Hann window respectively. Panels (c) and (d) show the QSF averaged over $N_{\rm{s}}=15$ disorder realizations, again without (c) and with (d) the Hann window respectively, showing the effect of each processing step on the final result.}
\label{fig.num}
\end{figure}
\end{center}

In this Appendix, we demonstrate the processing steps used to obtain the QSF shown in the main text (Fig.~\ref{fig.qsf_random}). There are two main steps which we employ. The first is the application of a suitable window function to the data obtained from the time-dependent variational principle, before taking the Fourier transform: this helps to reduce numerical artefacts due to boundary effects. Here, we use the Hann window function, a common choice in spectral analysis, though we have verified that different choices of window function give qualitatively similar results. The second step is a disorder average, which smooths out sample-to-sample fluctuations.

In Fig.~\ref{fig.num}, we show four examples of the QSF of $g_1(x,t)$ at a fixed value of $U=7.5$, $\mu/U = 0.15$. In panel (a), we show a single disorder realization, and do not apply a window function. The resulting data is noisy, and displays a strong peak at $k=0$. Panel (b) is the same but with the window function applied. The result is a clearer signal, but still visibly noisy. In panel (c), we show the QSF averaged over $N_{\rm{s}}=15$ disorder realizations, still without the window function. The resulting data is smoother due to the disorder average, but still displays a strong $k=0$ peak. In panel (d), we show the method used in the main text [shown in Fig.~\ref{fig.qsf_random}(a)], where we apply a window function \emph{and} a disorder average: the resulting data is broadened slightly due to the use of the window function, but is smoother than panels (a) and (b), and displays a much stronger signal than in panel (c). The strong peak at $k=0$ has gone, enabling us to more clearly see the underlying structure of the QSF. 

\section{Divergences associated to energy transitions in the strongly interacting SF}
\label{app:calcul_4J_sink/2}
In this Appendix, we discuss the origin of the signal observed in the QSF of the density in the strongly interacting SF phase in the presence of weak disorder. 
To understand the transitions probed by the density, it should be noted that the excitations in this regime are non-local collective excitations (phonons) which we would not expect to be able to excite with a purely local observable, and as such the density cannot be used to probe the excitation spectrum. Instead, after a weak global quench the density probes the transition energies $\omega\simeq E_{n'}-E_{n}$ where $n$ and $n'$ label two different low-energy eigenstates belonging to the same energy manifold. We assume weak disorder and we discard the shift of the energy resonances $V_{n'}-V_{n}$. The main role of the weak disorder here is to break translation invariance of the system and therefore prevent the emergence of the additional momentum selection rule imposing $P_{n'}=P_{n}$ in the clean case. We also discard the broadening effects imposed by the disorder in momentum-space such that we consider the momentum selection rule $\delta(P_{n'}-P_{n'}-k)$ [see Eq.~\eqref{eq:QSF_to_comment_physical_effects}]. Then, we may write $P_{n'}=q$ and $P_{n}=q-k$ where $q$ is an arbitrary quasimomentum and rewrite the QSF as
\begin{equation}
\begin{split}
G(k,\omega)&\simeq \int\d q\,C_{q,q-k}\,\delta\left(E_{q}-E_{q-k}-\omega\right),\\
&\simeq \int \d q\,C_{q^{\star},q^{\star}-k}\,\left|\partial_{q}g_{k}(q^{\star}_{k}(\omega))\right|^{-1},
\end{split}
\end{equation}
where we assumed that $\forall q^{\star}_{k}(\omega)$,$\left.\partial_{q}g_{k}(q)\right|_{q=q^{\star}_{k}(\omega)}\neq 0$ with $q^{\star}_{k}(\omega)$ the $\omega$-dependent zeros of the function $g_{k}$ and we have defined for convenience
\begin{equation}
  \begin{split}
  &C_{q,q-k}=\rhoi^{q;q-k}\,\bra{q-k}\hat{O}\ket{q},\\
  &g_{k}(q,\omega)=E_{q}-E_{q-k}-\omega.
  \end{split}
  \end{equation}  
We have argued in the main text than the excitation spectrum in the strongly interacting regime was well-approximated by the following form
\begin{equation} 
 E_{k}=4\bef J\sin\left(\tfrac{k}{2}\right)\sin\left(\tfrac{k}{2}+k_{\rm{h}}\right),
 \end{equation}
where $k_{\rm{h}}$ is a parameter which spans the range $[-\kf;\kf]$. 
First, using trigonometric identities, the function $g_{k}$ may be rewritten in an exact way as
\begin{equation}
g_{k}(q,\omega)=4\bef J \sin\left(\tfrac{k}{2}\right)\sin\left(q-\tfrac{k}{2}+k_{\rm{h}}\right)-\omega,
\end{equation}
  so the zeros of $g_{k}(q,\omega)$ are then given by
  \begin{equation}
q^{\star}_{k}(\omega)-\tfrac{k}{2}+k_{\rm{h}}=\arcsin\left(\frac{\omega}{4\bef J\sin\left(\tfrac{k}{2}\right)}\right).
\end{equation}
Second, we evaluate the derivative of the argument of the energy selection rule at the points where it is zero
  \begin{equation}
  \begin{split}
\left.\partial_{q}g_{k}(q)\right|_{q=q^{\star}_{k}(\omega)}
&=4\bef J\sin\left(\tfrac{k}{2}\right)\cos(q^{\star}_{k}(\omega)-\tfrac{k}{2}+k_{\rm{h}})\\
&=4\bef J\sin\left(\tfrac{k}{2}\right)\sqrt{1-\left(\dfrac{\omega}{4\bef J\sin\left(\tfrac{k}{2}\right)}\right)^{2}}.
\end{split}
  \end{equation}
Therefore we expect the QSF of the density to display algebraic divergences coming from the energy differences, along the lines
  \begin{equation}
   \omega = \pm 4\bef J\sin\left(\tfrac{k}{2}\right).
   \end{equation}
This prediction yields an excellent agreement with the numerical results, as shown in Fig.~\ref{fig.qsf_random} and Fig.~\ref{fig.qsf_rb2}.

\vspace{1cm}

\bibliographystyle{apsrev4-2}
\bibliography{refs,refs2,biblioLSP}

\end{document}

%% file: journalsVaps.tex
\newcommand{\Jprl}{Phys. Rev. Lett.}

\newcommand{\Jpra}{Phys. Rev. A}
\newcommand{\Jprb}{Phys. Rev. B}

\newcommand{\Jepjd}{Eur. Phys. J. D}